%% file: main.tex
\documentclass[conference]{IEEEtran}
\IEEEoverridecommandlockouts
\usepackage{cite}
\usepackage{amsmath,amssymb,amsfonts}
\usepackage{algorithmic}
\usepackage{graphicx,color}
\usepackage{textcomp}

\usepackage{ulem}
\usepackage{hyperref}
\usepackage{xcolor}
\usepackage{tabularx}
\usepackage{booktabs}
\usepackage[table]{xcolor}
\usepackage{tikz}
\usetikzlibrary{shapes.geometric, arrows.meta, positioning, fit, backgrounds, calc}

\usepackage{enumitem}

\usepackage[most]{tcolorbox} 
\usepackage{chngcntr} 

\def\BibTeX{{\rm B\kern-.05em{\sc i\kern-.025em b}\kern-.08em
    T\kern-.1667em\lower.7ex\hbox{E}\kern-.125emX}}
\AtBeginDocument{\definecolor{ojcolor}{cmyk}{0.93,0.59,0.15,0.02}}

\begin{document}

\title{Zero-Trust Strategies for O-RAN Cellular Networks:\\
Principles, Challenges and Research Directions}

\author{\IEEEauthorblockN{Charalampos Katsis}
\IEEEauthorblockA{\textit{Department of Computer Science} \\
\textit{Purdue University}\\
West Lafayette, Indiana, USA\\
ckatsis@purdue.edu}
\and
\IEEEauthorblockN{Imtiaz Karim}
\IEEEauthorblockA{\textit{Department of Computer Science} \\
\textit{The University of Texas at Dallas}\\
Richardson, Texas, USA\\
imtiaz.karim@utdallas.edu}
\and
\IEEEauthorblockN{Elisa Bertino}
\IEEEauthorblockA{\textit{Department of Computer Science} \\
\textit{Purdue University}\\
West Lafayette, Indiana, USA\\
bertino@purdue.edu}
}

\tcbset{
  colback=gray!3,
  colframe=black!20,
  coltitle=black,
  fonttitle=\bfseries,
  boxrule=0.4pt,
  arc=2pt,
  left=8pt,right=8pt,top=6pt,bottom=6pt,
}

\newtcolorbox{takeawaybox}[1][]{%
  enhanced, breakable,
  borderline west={2pt}{0pt}{blue!55},
  boxed title style={
    colback=gray!10,
    colframe=black!20,
    boxrule=0.4pt,
  },
  attach boxed title to top left={xshift=1mm,yshift*=-2mm},
  #1
}

\newcounter{takeaway}

\newcommand{\takeawaytitle}[1]{%
  \refstepcounter{takeaway}%
  Takeaway~\thetakeaway%
  \if\relax\detokenize{#1}\relax\else\ --- #1\fi
}
\newcommand{\ik}[1]{\textcolor{purple}{\textbf{Imtiaz: #1}}}
\newcommand{\ZTATakeaway}[2][]{%
  \begin{takeawaybox}[title=\takeawaytitle{#1}]%
  #2
  \end{takeawaybox}%
}

\newenvironment{ZTATakeaways}[1][]{%
  \begin{takeawaybox}[title=\takeawaytitle{#1}]%
  \begin{itemize}[leftmargin=*,itemsep=0.1em,topsep=0.1em]
}{%
  \end{itemize}%
  \end{takeawaybox}%
}

\maketitle

\begin{abstract}
Cellular networks are foundational to modern communication, supporting a broad range of applications, from civilian use to enterprise systems and military tactical networks. The advent of fifth-generation and beyond cellular networks (B5G) introduces emerging compute capabilities into the Radio Access Network (RAN), transforming it from a traditionally closed, vendor-locked infrastructure into an open and programmable ecosystem. This evolution, exemplified by Open-RAN (O-RAN), enables the deployment of control-plane applications from diverse sources, which can dynamically influence user-plane traffic in response to real-time events. As cellular infrastructures become more disaggregated and software-driven, security becomes an increasingly critical concern. Zero-Trust Architecture (ZTA) has emerged as a promising security paradigm that discards implicit trust assumptions by acknowledging that threats may arise from both external and internal sources. ZTA mandates comprehensive and fine-grained security mechanisms across both control and user planes to contain adversarial movements and enhance breach detection and attack response actions. In this paper, we explore the adoption of ZTA in the context of 5G and beyond, with a particular focus on O-RAN as an architectural enabler. We analyze how ZTA principles align with the architectural and operational characteristics of O-RAN, and identify key challenges and opportunities for embedding zero-trust mechanisms within O-RAN-based cellular networks.
\end{abstract}

\begin{IEEEkeywords}
ZTA, O-RAN, Cellular Networks, 5G, B5G
\end{IEEEkeywords}

\maketitle

\input{sections/introduction}

\input{sections/related_work}

\input{sections/idm_ac}
\input{sections/zta}

\input{sections/oran}

\input{sections/control_plane_zta}
\input{sections/user_plane_zta}

\input{sections/industry}
\input{sections/conclusion}

\section*{Acknowledgment}
This work was supported in part by NSF under grant 2112471, the University of Texas System Rising STARs Award (No. 40071109), and the startup funding from the University of Texas at Dallas.

\bibliographystyle{IEEEtran}
\bibliography{literature-base}

\end{document}

%% file: sections/introduction.tex
\section{\MakeUppercase{Introduction}}

The Open Radio Access Network (O-RAN) paradigm marks a substantive shift in mobile network architecture: the Radio Access Network (RAN) becomes open, interoperable, virtualized, and intelligent, with compute pushed toward the edge where user-plane devices attach. Core RAN building blocks radio units (O-RUs), distributed and centralized units (O-DUs/O-CUs), and the intelligent control layers (Near-RT/Non-RT RIC and the SMO) can be implemented or procured from different parties. Even controller-resident software modules (e.g., xApps/rApps) may be provided different vendors. Communication across these components is enabled by open interfaces specified by the O-RAN Alliance~\cite{O-RAN_ALLIANCE_2025}. Many elements are cloud-native and virtualized, allowing deployment on general-purpose cloud infrastructure and, in some cases, shared underlying hardware across multiple network operators. This disaggregated architecture embraces openness, interoperability, and virtualization and promises a more flexible and innovative ecosystem for 5G and beyond. It also increases operator visibility (e.g., per-cell and per-UE (User Equipment) performance indicators, traffic KPIs) and control (e.g., network slicing, device admission policies, ML-assisted optimization for operations, mobility, and handovers).

However, the same disaggregation expands and complicates the attack surface. Reliance on cloud-native components, third-party software, and open interfaces raises fundamental questions about trust establishment, authentication, attestation, and policy enforcement across a multi-vendor, distributed infrastructure~\cite{xing2024criticality, groen2024timesafe, lin20255g}. On the contrary, the Zero-Trust Architecture (ZTA) has emerged as a modern security principle ``never trust, always verify'' emphasizing continuous verification, least privilege, and identity-centric, posture-aware policies. While ZTA is widely applied in enterprise and cloud systems, O-RAN’s real-time constraints, heterogeneous interfaces (e.g., E2, A1, O1/O2, Open Fronthaul), and multi-stakeholder procurement make a direct transplant insufficient; practical ZT for O-RAN must be RAN-aware and enforceable at RAN timescales.

Despite the intuitive alignment, ZTA’s application within O-RAN remains underexplored. The Third Generation Partnership Project (3GPP)~\cite{3GPP_2025}, the standards body responsible for defining 4G and 5G system architectures along with the O-RAN Alliance, provides important baseline protections such as secure transport, mutual authentication, and integrity mechanisms. However, these frameworks primarily address component-level security and lack a holistic, dynamic, and verifiable trust model that adapts to evolving threats and the multi-stakeholder nature of O-RAN deployments.
In particular, O-RAN’s performance requirements and deployment realities call for tailored ZT mechanisms that go beyond traditional enterprise ZTA deployments.

\subsection{\MakeUppercase{Paper scope}} 

This paper examines the architectural and systems-level challenges of embedding ZT into O-RAN–based cellular infrastructures. We first map ZTA’s foundational elements as formalized by the U.S. National Institute of Standards and Technology (NIST)~\cite{stafford2020zero} to the control and data planes of disaggregated 5G/O-RAN deployments. We then analyze security challenges in the O-RAN control plane and assess how they are acknowledged and partially addressed by the O-RAN Alliance. Beyond the current guidance, we surface gaps that fall squarely under ZTA requirements, for example the procurement and onboarding processes for external components (xApps/rApps/models) and the precise scoping of what those components are allowed to do within the O-RAN environment. Crucially, ZT cannot stop at the control plane: the user plane must also enforce explicit, mission-scoped authorization conditioned on entity ``health'' and context. To this end, we propose a policy-oriented framework that leverages programmable RAN controllers (e.g., Near-RT RIC) to enforce fine-grained, stateful, and adaptive policies, coordinated by the SMO layer. Our analysis highlights immediate systems opportunities and longer-term architectural directions toward trustworthy, programmable, and resilient cellular networks.

\subsection{\MakeUppercase{Contributions}} 

While prior literature has explored ZTA within the O-RAN control plane, to the best of our knowledge, this is the first work that provides a holistic ZTA translation across both the O-RAN control and user planes. We uniquely address structural security gaps within the control plane ecosystem, formulate data-plane enforcement mechanisms, and discuss implications across diverse administrative trust boundaries. 
The key contributions of this paper are as follows:

\begin{itemize}[leftmargin=*, itemsep=0.1em]
\item We present a comprehensive discussion of core ZTA security controls, emphasizing identity management and access control mechanisms.

\item We review major ZTA guidelines and recommendations, including those 
by NIST and the Cybersecurity and Infrastructure Security Agency (CISA).

\item We analyze relevant O-RAN technical specifications to assess the current level of ZTA adoption and identify existing security gaps. Furthermore, we survey related academic work and highlight open research challenges in this domain.

\item We introduce a new research direction focused on extending ZTA principles to the O-RAN data plane by leveraging its architectural and functional capabilities. In doing so, we propose a policy framework for authoring and managing fine-grained, stateful, and adaptive policies.

\item We examine industrial perspectives by analyzing ZT initiatives and architectural approaches proposed by leading vendors such as Ericsson and Nokia.
\end{itemize}

\begin{table*}[t]
\centering
\scriptsize
\setlength{\tabcolsep}{3pt}
\renewcommand{\arraystretch}{0.9}
\caption{Abbreviations}
\label{tab:abbrev}
\begin{tabularx}{\textwidth}{@{}lX lX lX lX@{}}
\toprule
\textbf{Abbrev.} & \textbf{Definition} & \textbf{Abbrev.} & \textbf{Definition} & \textbf{Abbrev.} & \textbf{Definition} & \textbf{Abbrev.} & \textbf{Definition} \\
\midrule
3GPP & 3rd Generation Partnership Project & ABAC & Attribute-Based Access Control & AC & Access Control & ACC & Access Control Contract \\
AKA & Authentication and Key Agreement & AMF & Access and Mobility Management Function & API & Application Programming Interface & ASIC & Application-Specific Integrated Circuit \\
B5G & Fifth-generation and beyond cellular networks & BSR & Buffer Status Report & CDM & Continuous Diagnostics and Mitigation & CI/CD & Continuous Integration / Continuous Delivery \\
CISA & Cybersecurity and Infrastructure Security Agency & CQI & Channel Quality Indicator & CSI & Channel State Information & CSP & Communications Service Provider \\
CU & Centralized (Control) Unit & CU-CP & CU Control Plane & CU-UP & CU User Plane & CVE & Common Vulnerabilities and Exposures \\
DoS & Denial-of-Service & DRB & Data Radio Bearer & DU & Distributed Unit & E2AP & E2 Application Protocol \\
EI & Enrichment Information & EIAP & Ericsson Intelligent Automation Platform & eMBB & Enhanced Mobile Broadband & ESM & Ericsson Security Manager \\
F1AP & F1 Application Protocol & FPGA & Field-Programmable Gate Array & GNN & Graph Neural Network & HSM & Hardware Security Module \\
IM & Identity Management & IMS & Identity Management System & IMSI & International Mobile Subscriber Identity & INSSA & Intelligent Network Security State Analysis \\
IoT & Internet of Things & KPI & Key Performance Indicator & KPM & Key Performance Measurement & LLM & Large Language Model \\
LSTM & Long Short-Term Memory & MAC & Medium Access Control & MEC & Multi-access Edge Computing & ML & Machine Learning \\
MNO & Mobile Network Operator & mTLS & Mutual TLS & NAC & Network Access Control & NAS & Non-Access Stratum \\
Near-RT RIC & Near-Real-Time RAN Intelligent Controllers & NF & Network Function & NIST & National Institute of Standards and Technology & Non-RT RIC & Non-Real-Time RAN Intelligent Controllers \\
NR & New Radio & OFH & Open Fronthaul & O-RAN & Open Radio Access Network & OTP & One-Time Password \\
OWASP & Open Web Application Security Project & PA & Policy Administrator & PCF & Policy Control Function & PDCP & Packet Data Convergence Protocol \\
PDP & Policy Decision Point & PDU & Protocol Data Unit & PE & Policy Engine & PEP & Policy Enforcement Point \\
PHY & Physical layer & PKI & Public Key Infrastructure & PM & Performance Measurement & PRB & Physical Resource Block \\
QoS & Quality of Service & RAdAC & Risk-Adaptive Access Control & RAN & Radio Access Network & RBAC & Role-Based Access Control \\
RF & Radio Frequency & RLC & Radio Link Control & RRC & Radio Resource Control & RSRP & Reference Signal Received Power \\
RSRQ & Reference Signal Received Quality & RT-RIC & Real-Time RIC & RU & Radio Unit & SBOM & Software Bill of Materials \\
SCADA & Supervisory Control and Data Acquisition & SDAP & Service Data Adaptation Protocol & SDN & Software-Defined Networking & SECSM & Security Service Module \\
SIEM & Security Information and Event Management & SIM & Subscriber Identity Module & SINR & Signal-to-Interference-plus-Noise Ratio & SLA & Service-Level Agreement \\
SMF & Session Management Function & SMO & Service Management and Orchestration & S-NSSAI & Single Network Slice Selection Assistance Information & SRM & Ericsson’s Security Reliability Model \\
SUPI & Subscription Permanent Identifier & TLS & Transport Layer Security & UAV & Unmanned Aerial Vehicle & UDM & Unified Data Management \\
UE & User Equipment & UPF & User Plane Function & URLLC & Ultra-Reliable Low Latency Communications & VNF & Virtual Network Function \\
XRF & xApp Repository Function & ZTA & Zero-Trust Architecture & ZTMM & Zero-Trust Maturity Model (CISA) & & \\
\bottomrule
\end{tabularx}
\end{table*}

\subsection{\MakeUppercase{Paper Structure}}

\begin{figure}[htbp]
\centerline{\includegraphics[width=\columnwidth]{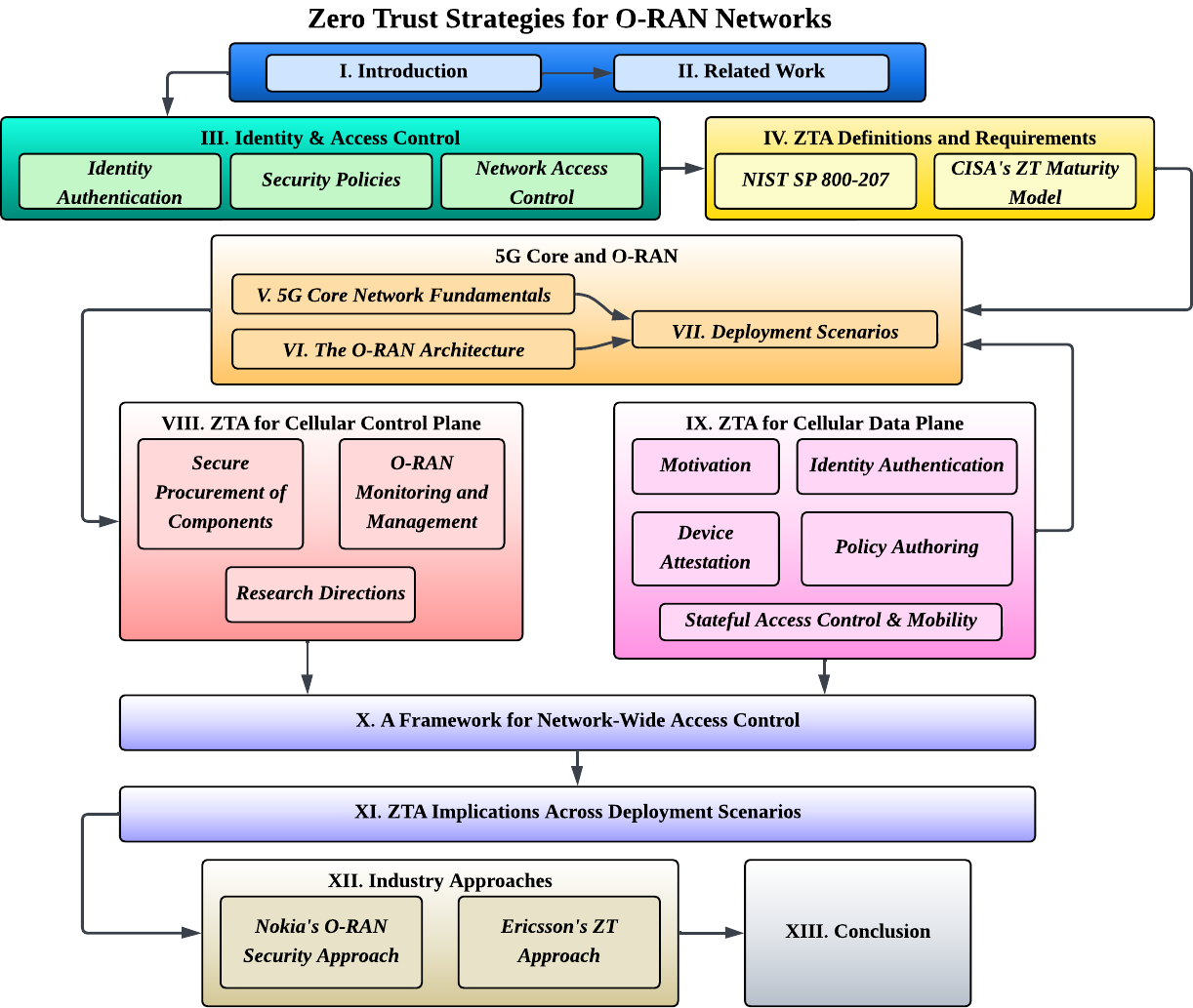}}
\caption{\label{fig:paper_map} Paper structure.
}
\end{figure}

The remainder of this paper is organized as follows (see Figure~\ref{fig:paper_map}).
Section~\ref{sec:2relatedwork} discusses related surveys on O-RAN security.
Section~\ref{sec:ID-management} outlines the fundamentals of Identity Management and Access Control, covering authentication, access control models, and least-privilege enforcement as foundational ZT controls.
Section~\ref{sec:zta} overviews the ZTA and key standards NIST SP~800-207 and the CISA Zero-Trust Maturity Model that inform secure, policy-driven system design.
Section~\ref{sec:5g_foundations} summarizes the 5G Core architecture, while Section~\ref{sec:oran} details the O-RAN framework, emphasizing its open interfaces, control hierarchy, and RAN-to-core communication workflows.
Section~\ref{sec:oran-deployment-scenarios} analyzes operator, private, hybrid, and multi-operator shared deployments to contextualize ZTA adoption.
Section~\ref{sec:cp-zta} applies ZTA principles to the control plane, focusing on secure procurement, continuous monitoring, and lifecycle management of O-RAN components.
Section~\ref{sec:up-zta} extends the analysis to the user (data) plane, discussing a motivating scenario and mechanisms such as device attestation, identity-aware state transfers, and adaptive network access control.
Section~\ref{sec:proposed-framework} introduces a framework for network-wide policy control, describing policy specification and enforcement through programmable RAN controllers and SMO coordination.
Section~\ref{sec:deployment-implications} discusses the implications of ZTA across various O-RAN deployment scenarios
Section~\ref{sec:industry} reviews industry efforts particularly by Nokia and Ericsson aligning O-RAN security with ZTA principles, followed by a discussion of open research challenges.
Finally, Section~\ref{sec:conclusion} concludes the paper.


%% file: sections/related_work.tex
\section{\MakeUppercase{Related Surveys for O-RAN Security}}
\label{sec:2relatedwork}

\begin{table*}[t]
\centering
\caption{Comparison of our work with existing literature.}
\label{tab:comparison}
\renewcommand{\arraystretch}{1.2}
\begin{tabularx}{\textwidth}{@{} X c c c c c c c c c c c @{}}
\toprule
\textbf{Related Works} & \textbf{\cite{mehrban2025integrating}} & \textbf{\cite{park2024investigation}} & \textbf{\cite{agarwal2025open}} & \textbf{\cite{hoffmann2023open}} & \textbf{\cite{wani2024open}} & \textbf{\cite{soltani2025intelligent}} & \textbf{~\cite{amachaghi2024survey}} & \textbf{\cite{bonati2020open}} & \textbf{\cite{ramezanpour2022intelligent}} & \textbf{\cite{polese2023understanding} } & \textbf{Our Work} \\
\midrule
\textbf{Paper Type} & S & R & S & S & S & S & S & S & P & S\&T & S\&P \\
\midrule
ZTA-Specific Threat Landscape & \checkmark & $\times$ & \checkmark & $\times$ & $\times$ & $\times$ & $\times$ & $\times$ & \checkmark & $\circ$ & \checkmark \\
O-RAN Deployment Scenarios & $\times$ & $\times$ & $\times$ & $\times$ & $\times$ & $\times$ & $\times$ & $\times$ & $\times$ & $\circ$ & \checkmark \\
Secure Procurement of Third-Party Components (xApps/rApps/dApps) & $\circ$ & $\times$ & $\circ$ & $\times$ & $\times$ & $\circ$ & $\times$ & $\times$ & $\times$ & $\circ$ & \checkmark \\
Secure AI/ML Procurement & $\times$ & $\circ$ & $\times$ & $\times$ & $\times$ & $\times$ & $\times$ & $\times$ & $\times$ & $\circ$ & \checkmark \\
Secure RAN Component Procurement \& Resource Sharing & \checkmark & $\times$ & $\times$ & $\times$ & $\times$ & $\times$ & $\times$ & $\times$ & $\times$ & $\circ$ & \checkmark \\
O-RAN Monitoring, Management, \& Task Scoping & $\times$ & $\circ$ & $\circ$ & $\circ$ & $\circ$ & $\circ$ & $\circ$ & $\times$ & $\circ$ & $\times$ & \checkmark \\
Control Plane ZTA Research Directions & \checkmark & $\times$ & $\circ$ & $\times$ & $\times$ & \checkmark & \checkmark & $\times$ & \checkmark & $\times$ & \checkmark \\
Application of ZTA on the Cellular Data Plane & $\times$ & $\times$ & $\times$ & $\times$ & $\times$ & $\times$ & $\times$ & $\times$ & $\circ$ & $\times$ & \checkmark \\
Data Plane ZTA Motivating Scenarios & $\times$ & $\circ$ & $\circ$ & $\times$ & $\circ$ & $\times$ & $\circ$ & $\times$ & $\times$ & $\times$ & \checkmark \\
Data Plane ZTA Mechanisms (Attestation, Identity, Stateful Policies) & $\times$ & $\times$ & $\times$ & $\times$ & $\times$ & $\times$ & $\circ$ & $\times$ & $\circ$ & $\times$ & \checkmark \\
ZTA Approaches by Industry & $\times$ & $\times$ & $\times$ & $\times$ & $\times$ & $\times$ & \checkmark & $\times$ & $\times$ & $\times$ & \checkmark \\
\bottomrule
\multicolumn{12}{@{}p{\textwidth}@{}}{\footnotesize \textbf{Legend:} \checkmark: Yes, \quad $\times$: No, \quad $\circ$: Partial. \quad \textbf{Types:} S: Survey, R: Review, P: Position, T: Tutorial.} \\
\end{tabularx}
\end{table*}

In this section, we review the existing survey, position and review papers on O-RAN security and ZTA, highlighting the specific gaps our work addresses. A detailed comparison of our contributions against these existing studies is summarized in Table~\ref{tab:comparison}.

Mehrban et al.~\cite{mehrban2025integrating} provide a comprehensive survey focusing on the integration of Zero Trust Architecture (ZTA) into O-RAN. Their work systematically maps core ZTA components namely the Policy Engine (PE), Policy Administrator (PA), and Policy Enforcement Points (PEPs) to the specific logical layers of the O-RAN architecture, such as the Non-RT and Near-RT RICs. Furthermore, they highlight several proposed state-of-the-art implementations, including Risk-Adaptive Access Control (RAdAC) mechanisms to dynamically modulate verification depth, post-quantum cryptographic defenses for the open fronthaul, and blockchain-enabled decentralized identity management. Their analysis, however, remains primarily centered on control and management plane vulnerabilities, such as insecure communication interfaces and high-level third-party procurement issues. Crucially, their survey does not fully extend the ZTA tenets defined in the NIST 800-207 specification~\cite{stafford2020zero} to the cellular data plane, leaving critical aspects like user-plane traffic inspection, device-level identity authentication, and stateful policy enforcement largely unaddressed. Furthermore, their survey does not structurally analyze how diverse administrative boundaries impact ZTA across varying O-RAN deployment scenarios such as private cellular networks, hybrid public-private setups, and multi-operator shared RAN environments utilizing Shared O-RUs. Moreover, their work lacks a granular analysis of current O-RAN Alliance use cases, specifically regarding the functional control actions and parameters monitored by applications (e.g., xApps and rApps) to dynamically dictate RAN behavior. Finally, their analysis relies strictly on standards and academic literature, omitting applied industrial ZTA proposals and solutions by leading telecommunications vendors.

Park et al.~\cite{park2024investigation} investigate O-RAN specifications, detailing the general architecture, operational use cases (such as V2X handover management, UAV resource allocation, and traffic steering), and standard security requirements. Their work systematically categorizes conventional vulnerabilities—including spoofing, sniffing, jamming, and component-specific threats—across the disaggregated O-RAN components and interfaces. Furthermore, they outline standard mitigation strategies and future research directions, such as leveraging Graph Neural Networks (GNNs), automated reinforcement learning (AutoRL), and federated learning to enhance network security and adaptability. Their analysis, however, approaches network defense only from a conventional communication security standpoint such as the use of standard security mechanisms such as IPsec, TLS, and OAuth 2.0 authentication but they do not discuss the several issues that ZTA introduces.  
Furthermore, although they discuss data-plane motivating scenarios like UAVs and V2X for operational resource allocation and mobility, they do not explore the cellular user plane from a ZTA security standpoint. Crucially, their work lacks an analysis of low-level data-plane ZT mechanisms, such as remote device attestation, continuous identity verification, and network access control. Finally, their architectural analysis relies on a single, generic deployment model, entirely omitting the structural security nuances of diverse administrative boundaries (such as private networks, hybrid setups, or multi-operator shared environments) and their review is confined to academic solutions, lacking an evaluation of applied ZTA frameworks from commercial telecommunications vendors. Finally, the paper does not discuss ZTA proposals and solutions by leading telecommunications vendors.

Agarwal et al.~\cite{agarwal2025open} provide a survey on the evolution of O-RAN architectures to support 6G networks, heavily emphasizing performance metrics such as energy efficiency, ultra-low latency, and the integration of advanced technologies like Digital Twins, Massive MIMO, and V2X communications. In their security analysis, the authors discuss the importance of ZT principles to address the expanded attack surface of the multi-vendor O-RAN ecosystem. They map ZTA tenets, such as micro-segmentation, continuous verification, mutual authentication, and Role-Based Access Control (RBAC), to O-RAN interfaces and components. Furthermore, they highlight the necessity of robust CI/CD pipelines to manage third-party applications and future quantum-resistant cryptography to defend against emerging threats. 
While they note the threats of malicious xApps and adversarial attacks against embedded AI/ML models, they focus on CI/CD lifecycle management rather than secure procurement processes; they omit requirements for continuous build reproducibility, benchmark datasets, training metadata, or model watermarking to verify opaque models. Similarly, while acknowledging external hardware threats, such as rogue base stations, they do not analyze the structural isolation and supply chain risks of Shared O-RUs in multi-tenant environments. Furthermore, their security analysis remains confined to the control and management planes, entirely overlooking the application of ZTA to the cellular user plane and leaving critical data-plane mechanisms, like remote device attestation and stateful network access control, unaddressed. Finally, their architectural review categorizes deployment scenarios strictly by the physical and logical placement of cloud resources to optimize latency, rather than structurally analyzing how diverse administrative and trust boundaries (such as private, hybrid, or multi-operator shared environments) impact multi-tenant policy enforcement, nor do they evaluate applied commercial ZTA frameworks.

Hoffmann et al.~\cite{hoffmann2023open} focus on the evaluation of xApp design and deployment, highlighting the lessons learned from integrating third-party applications across several open-source RAN Intelligent Controller (RIC) platforms, such as OAI, OSC, and ONF. The authors detail the implementation of specific use cases, including beam mobility management, signaling storm detection, traffic steering, and Quality of Service (QoS), based resource allocation—to expose current architectural ambiguities and standardization gaps. In their security discussion, they identify the expanded attack surface introduced by O-RAN, specifically noting vulnerabilities within open interfaces and the susceptibility of edge-deployed AI/ML algorithms to poisoning, evasion, and inference attacks. Furthermore, they highlight the critical operational needs for intelligent conflict management to arbitrate overlapping xApp control actions, as well as the necessity for automated testing and continuous delivery pipelines to manage third-party software integration. However, their analysis does not adopt or mention the ZTA framework, nor does it propose structural ZTA enforcement mechanisms. While they strongly advocate for automated xApp delivery and testing, they approach this strictly from a continuous integration and software engineering perspective rather than a security validation standpoint. Consequently, they do not introduce a secure procurement discussions utilizing cryptographic validation, continuous build reproducibility, or benchmark datasets to verify opaque AI models before deployment. Furthermore, their analysis remains strictly focused on control-plane implementation challenges within open-source platforms, leaving the application of stateful ZTA policies on the cellular data plane, the structural impact of administrative trust boundaries (e.g., hybrid or multi-operator deployments), and the evaluation of commercial industrial ZTA frameworks entirely unaddressed.

Wani et al.~\cite{wani2024open} provide an overview of the O-RAN ecosystem, detailing its evolutionary trajectory from traditional monolithic base stations to disaggregated, cloud-native architectures. The authors introduce the core O-RAN components, standardized open interfaces, and operational use cases such as V2X dynamic handover and energy-saving management. In their security analysis, they summarize the expanded threat landscape introduced by O-RAN, relying primarily on O-RAN Alliance Working Group 11 threat models. They outline vulnerabilities such as malicious or conflicting xApps and rApps, hardware supply chain backdoors, and the susceptibility of embedded ML models to evasion and poisoning attacks.
While they note that O-RAN’s overarching security goal is to implement a robust ZT model, their work serves primarily as an introductory survey rather than a structural security proposal. Consequently, the authors do not propose or evaluate specific ZTA enforcement mechanisms. 
Furthermore, their analysis does not extend ZTA principles to the cellular user plane, omitting mechanisms like remote device attestation and stateful network access control. Finally, their work does not evaluate the structural impact of administrative trust boundaries—such as private, hybrid, or multi-operator shared environments nor does it assess applied commercial ZTA implementations.

Soltani et al.~\cite{soltani2025intelligent} analyze the security implications of intelligent control in 6G-ready O-RAN, focusing on how the RAN Intelligent Controller (RIC) expands the network's attack surface. The authors map specific vulnerabilities, such as API exploitation, uncertified radio resource access, and conflicting radio access policies, to the Near-RT RIC. Furthermore, they investigate vulnerabilities inherited from Software-Defined Networking (SDN) controllers, such as link layer discovery protocol (LLDP) manipulation and topology poisoning. To mitigate these risks, the survey discusses current O-RAN working group 11 (WG11) recommendations, including secure xApp onboarding via digital signatures, OAuth 2.0 for API access, and Role-Based Access Control (RBAC) for database queries.
However, while the authors reference ZTA as a promising framework to suppress hazards, their work serves as a targeted security analysis rather than a structural architecture proposal; they do not explicitly frame the threat landscape around core ZTA principles or map specific ZTA tenets to vulnerabilities. Regarding third-party procurement, they discuss secure onboarding and registration exclusively for xApps, relying on static certification and signature validation, but they entirely omit the secure procurement frameworks required for rApps and dApps, nor do they demand continuous build reproducibility. Similarly, while they identify the risk of poisoned AI/ML models, they do not introduce procurement requirements such as benchmark datasets or training metadata to verify opaque models prior to deployment. Furthermore, their analysis remains strictly confined to the control plane and RIC intelligence layers, leaving the applications on the cellular user plane  entirely unaddressed. Finally, their work does not evaluate the structural impact of administrative trust boundaries across different deployments or assess applied commercial ZTA implementations.

Amachaghi et al.~\cite{amachaghi2024survey} present a survey focusing on Intrusion Detection Systems (IDS) within O-RAN environments. The authors classify O-RAN security challenges into technical threats—such as infrastructure vulnerabilities, application attacks, network interception, and access control issues—and non-technical threats, including physical security and supply chain risks. Furthermore, they evaluate real-world case studies, specifically detailing Rakuten Mobile's implementation of a Zero Trust Network (ZTN) architecture, Identity and Access Management (IAM), secure API transactions, and secure DevOps practices. To mitigate emerging threats, the survey proposes extensive operational and control-plane security research directions, including the application of Moving Target Defense (MTD) against AI/ML poisoning, blockchain for decentralized authentication, and Large Language Models (LLMs) for intent-driven security policies.
However, while Amachaghi et al. thoroughly explore operational and AI-driven security research, their work serves primarily as a survey of IDS mechanisms rather than a structural architecture proposal; they do not explicitly frame their threat landscape around core ZTA tenets. Regarding third-party integration, they do not introduce a secure procurement challenges for xApps, rApps, or dApps utilizing continuous build reproducibility, nor do they establish procurement requirements—such as benchmark datasets or training metadata—to verify opaque AI/ML models before deployment. Furthermore, their analysis does not structurally examine multi-tenant isolation risks for Shared O-RUs, nor do they categorize deployments by administrative trust boundaries to illustrate shifting security perimeters. Finally, their data-plane focus remains strictly on monitoring and intrusion detection, leaving the application of network access control policies on the cellular user plane, as well as critical mechanisms like remote device attestation and stateful network access control, entirely unaddressed.

Bonati et al.~\cite{bonati2020open} provide a comprehensive survey on open-source, programmable, and virtualized software frameworks for 5G cellular networks. The authors detail a full-stack ecosystem, including implementations of the Radio Access Network (e.g., OpenAirInterface, srsLTE), core network gateways, virtualization orchestrators (e.g., ONAP, OSM), and multi-access edge computing (MEC) architectures. In assessing the operational roadblocks for 5G deployments, the survey highlights critical open security research directions; specifically, the authors emphasize that the exposure of Near-RT RIC APIs to third-party applications (xApps) introduces new vulnerabilities, necessitating rigorous security-by-design principles and community auditing for open-source cellular software.
However, while Bonati et al. exhaustively categorize open-source infrastructure capabilities and call for general software security research, their work serves as an architectural software compendium rather than a structural security proposal; they do not propose Control Plane ZTA-specific research directions, nor do they explicitly frame a threat landscape around core ZTA tenets. Regarding third-party integration, while they note the risks of open APIs, they do not discuss secure procurement approaches for xApps, rApps, or dApps, nor do they establish procurement requirements to verify opaque AI/ML models before deployment. Furthermore, their analysis does not structurally examine multi-tenant isolation risks for shared hardware from a ZTA perspective, nor do they propose granular task-scoping mechanisms using machine-readable O-RAN specifications. Finally, the application of stateful ZTA policies on the cellular user plane, remote device attestation, and data-plane network access control remain entirely unaddressed.

Ramezanpour et al.~\cite{ramezanpour2022intelligent} introduce an Intelligent Zero Trust Architecture (i-ZTA) for 5G/6G networks, explicitly framing their threat landscape around NIST and DoD Zero Trust pillars to address perimeter dilution and lateral movement vulnerabilities. The authors propose a dynamic, AI-driven framework that integrates directly into the O-RAN architecture to secure both the control and data planes. Specifically, their architecture introduces an Intelligent Policy Engine (IPE) that utilizes reinforcement learning and Long Short-Term Memory (LSTM) networks to track historical user behavior and execute stateful access decisions. Furthermore, they propose an Intelligent Network Security State Analysis (INSSA) engine utilizing GNNs for holistic network monitoring and risk assessment alongside a Continuous Multi-Factor Authentication (CMFA) mechanism to persistently verify identity throughout an active data-plane session.
While Ramezanpour et al. provide a robust foundation for AI-driven continuous verification, their reliance on runtime monitoring leaves several structural ZTA requirements open for further exploration. Because their framework primarily evaluates trust dynamically during operation—such as verifying software image hashes at runtime—it does not explicitly address the pre-deployment secure procurement pipeline. Consequently, it lacks mechanisms to verify the continuous build reproducibility (which separates benign software updates and malicious updates) of third-party applications to prevent upstream supply chain attacks. Similarly, while the authors extensively utilize AI, they do not establish procurement requirements, such as benchmark datasets or model watermarking, to validate the integrity of the opaque AI models themselves before deployment. 
Furthermore, while their INSSA engine features strong anomaly detection, the architecture does not incorporate deterministic access control  mechanisms to strictly constrain application permissions by default.
Their architecture also makes critical assumptions regarding Core Network collaboration: it structurally relies on the 5G Core’s Network Exposure Function (NEF) for traffic steering to MEC hosts, the Access and Mobility Management Function (AMF) and the User Plane Function (UPF) for global session and traffic data , and the Unified Data Management (UDM) for authentication services. In hybrid or multi-operator deployments where an enterprise RAN operator does not possess administrative control over a public core network or lacks collaborative agreements with it, the proposed architecture will not hold. Finally, critical user-plane ZTA challenges remain unaddressed. For instance, while the architecture evaluates basic security states like software versions, it lacks robust remote device attestation protocols to verify UE firmware prior to access, as well as concrete mechanisms for continuous policy enforcement during inter-cell mobility.

Polese et al.~\cite{polese2023understanding} provide a survey detailing the architectural principles, interfaces, and algorithms enabling the O-RAN ecosystem. Their work highlights the extended threat surface introduced by O-RAN’s disaggregation and reliance on third-party components, systematically reviewing security challenges across the control and management planes. They identify key vulnerabilities arising from open-source code, supply chain tampering, and multi-tenant RAN sharing scenarios. Furthermore, they map the lifecycle of AI/ML workflows within the Near-RT and Non-RT RICs, specifically noting the susceptibility of these models to data poisoning and tampering attacks, and they acknowledge the O-RAN Alliance's broader goal of moving toward a zero-trust model.
While Polese et al. expertly identify the structural vulnerabilities driving the need for enhanced security in O-RAN, their work serves primarily as an observational survey of current specifications rather than an architectural proposal for ZTA enforcement. For instance, although they map out a threat landscape, they evaluate these vulnerabilities through a traditional cybersecurity lens rather than explicitly structuring them against formal ZTA frameworks (such as the CISA Zero Trust Maturity Model or NIST 800-207) to highlight failures in implicit trust. 
Finally, their security analysis remains heavily focused on the control plane and infrastructure interfaces. It leaves the cellular data plane unaddressed.

To address these critical gaps in the literature, our work provides a holistic examination of ZTA integration across the entire O-RAN ecosystem, extending core ZT tenets beyond the control and management planes directly to the cellular data plane. Within the control plane, we investigate structural vulnerabilities by establishing strict requirements for the secure procurement of third-party applications (xApps, rApps, dApps) and opaque AI/ML models, while outlining the necessity of granular task-scoping to regulate permissions across shared RAN nodes. Our work is the first to discuss ZTA-specific research directions to satisfy these requirements, including remote device attestation, continuous identity authentication, stateful network access control, and the synchronization of user equipment (UE) state transfers across distinct Near-RT RIC domains during mobility events. We synthesize these mechanisms into a comprehensive, policy-oriented framework that utilizes the Service Management and Orchestration (SMO) layer to author high-level security policies, leveraging programmable RAN controllers for fine-grained enforcement at runtime. Finally, we discuss ZTA strategies from leading telecommunications vendors, highlighting both the industry's progress in securing cloud-native RAN deployments and the remaining structural gaps that our proposed research directions aim to solve.

%% file: sections/idm_ac.tex
\section{\MakeUppercase{Identity Management and Access Control}}
\label{sec:ID-management}

In this section, we outline the fundamental concepts of Identity Management (IM) and Access Control (AC). While these mechanisms are well-established in traditional enterprise IT environments, understanding their core mechanics is a necessary prerequisite for adapting them to the highly distributed, multi-stakeholder ecosystem of O-RAN. Specifically, the principles of continuous identity authentication and least-privilege access control introduced here serve as the foundational building blocks for the ZTA strategies we propose in later sections.

Within interconnected environments, IM\footnote{Table~\ref{tab:abbrev} summarizes the abbreviations used throughout the paper.} oversees the administration of system actors, their digital profiles, and the credentials required to access resources~\cite{torres2012survey,katsis2025zero}. System actors, ranging from human personnel to autonomous software and IoT hardware, navigate the network utilizing distinct identities that correspond to their assigned operational roles. These identities enforce system accountability by dictating exactly which actions a profile can perform.

Before a system grants access, an entity must validate its claimed identity by presenting corresponding credentials. This authentication process can occur natively within the network or rely on a verified external provider. For instance, before executing a destructive operation like deleting a database table, a data engineer must supply valid credentials to confirm they possess the required privileges.

An entity’s digital lifecycle begins upon onboarding, where it is provisioned with a unique identity and initial credentials. This digital presence remains active as long as the organizational relationship continues. Once that relationship terminates, the system may permanently erase the identity or archive it strictly for compliance and auditing purposes. The credentials themselves, whether user-generated or provisioned by the enterprise, are dynamic. Administrators can force updates or temporarily suspend them (for instance, to freeze access during an active breach investigation) without completely terminating the user's account. Additionally, modern IM frameworks rely heavily on identity attributes to facilitate fine-grained access control. These attributes define security-relevant metadata, such as an employee's departmental role, a device's current attestation status, or its remote access capabilities.

To manage these lifecycles, enterprises typically deploy centralized, federated, or decentralized architectural models depending on their operational requirements~\cite{pohn2020overview,bertino2010}. Furthermore, while early IM systems predominantly focused on human users, contemporary solutions have expanded significantly to support the diverse identity management requirements of IoT infrastructures~\cite{won2018decentralized, singla2018blockchain}.

\subsection{\MakeUppercase{Identity Authentication}}

ZT architectures rely fundamentally on robust authentication mechanisms. Before accessing any resource, every entity (whether human, hardware, or software service) must definitively prove its identity. Organizations implement various authentication strategies depending on their security requirements, user experience goals, and technical capacity.

\noindent\textbf{Knowledge-Based Authentication.}
These mechanisms require users to recall a secret, such as a PIN or a password. While these remain the standard for securing web applications and personal devices, they frequently fall victim to brute-force attempts, phishing campaigns, and credential stuffing~\cite{lee2022password}. Some systems still ask security questions for account recovery, but attackers can easily guess or look up the answers online, making them completely inadequate as a primary security measure.

\noindent\textbf{Possession-Based Authentication.}
Users prove their identity by presenting a physical or digital item they own. Common examples include authenticator applications or One-Time Passwords (OTPs) routed through SMS. However, SMS routing leaves users open to SIM-swapping, and poorly implemented systems might generate predictable OTPs~\cite{siqi21}. Alternatively, organizations can issue digital certificates or physical hardware tokens (e.g., smart cards). These provide robust cryptographic security but often drive up deployment costs and require strict physical safeguarding to prevent theft~\cite{yusop2025advancing}.

\noindent\textbf{Inherence-Based Authentication.}
This approach verifies identities using unique biological or behavioral markers, ranging from iris patterns and fingerprints to typing speed and gait. Biometrics heavily resist traditional forgery attempts and deliver strong security guarantees. Even so, they introduce valid privacy concerns and occasionally frustrate users with false acceptance or rejection rates during the scanning process~\cite{eberz2017evaluating}.

To defend against the weaknesses of any single approach, modern security deployments heavily favor Multi-Factor Authentication (MFA), merging at least two distinct categories (such as requiring both a password and a hardware token). Furthermore, many environments now deploy continuous authentication. Instead of only checking credentials at login, these systems constantly analyze user activity streams to spot anomalies and verify the entity's identity throughout the entire session.

\subsection{\MakeUppercase{Access Control Policies}}
\label{sec:access_control_policies}

Security officers and network administrators establish Access Control (AC) policies to dictate exactly what operations entities can execute within a protected environment. We can broadly divide these directives into two categories: end-system controls and network-level controls.

\noindent\textbf{End System Access Control.}
These policies govern interactions with specific applications, databases or file systems. For instance, these rules dictate who can launch a particular software tool, write to a secure directory, or run SQL queries to alter database records. Managing these permissions becomes exceptionally difficult at scale, particularly when securing thousands of individual files or database tables. To overcome this hurdle, researchers have developed abstract policy models that simplify rule creation and streamline enforcement~\cite{jabal2019methods}. Within a ZTA, Role-Based Access Control (RBAC)~\cite{sandhu1998role} and Attribute-Based Access Control (ABAC)~\cite{hu2013guide} serve as the two most prominent models.

\noindent\textbf{Network Access Control (NAC).}
NAC policies govern endpoint communication by determining which devices can communicate with one another and what protocols they may use. Firewalls and similar enforcement points apply these policies by evaluating every passing packet against predefined rule sets. A standard rule contains two primary components: a traffic selector and a designated action. The traffic selector evaluates packet headers against specific criteria, such as source or destination IP addresses, port numbers, and protocol types. Depending on the hardware, advanced enforcement points might examine even deeper packet characteristics. If a packet satisfies every criterion in the selector, the system executes the corresponding action—such as forwarding the packet to its destination, mirroring it for analysis, or dropping it entirely. Because a single packet might trigger multiple conditions, administrators assign priority levels to each rule to ensure the system enforces the correct, highest-ranking outcome.

\subsection{\MakeUppercase{Least Privilege Access Control Policies}}

Operating under the principle of least privilege forms the backbone of ZTA~\cite{stafford2020zero, katsis2021can, katsis2024zt}. This model restricts an entity's permissions strictly to the bare minimum required to execute its assigned tasks. Rather than securing only the external perimeter, organizations must deploy and enforce these granular rules pervasively throughout the entire network environment.

Because a least-privilege framework operates on a strict ``default-deny'' posture, the system automatically blocks all access requests unless an explicit rule permits them. To prevent operational disruptions, administrators must design security policies that accurately and comprehensively map to the precise mission requirements of every network actor. By tightly constraining these rights, and immediately revoking them the moment they are no longer needed, this approach drastically reduces the attack surface and limits the potential for unauthorized lateral movement or insider threats~\cite{katsis2022neutron,katsis2021can,katsis2025NetSoft,katsis2024zt}.

%% file: sections/zta.tex
\section{\MakeUppercase{Zero-Trust Architecture}}
\label{sec:zta}

In this section, we review the foundational principles and maturity models of ZTA as defined by NIST and CISA. Although these frameworks were primarily designed for conventional enterprise IT and cloud networks, they provide the essential structural tenets, such as continuous verification, an assume-breach posture, and dynamic, risk-based access, required to secure next-generation cellular networks. We establish these standardized ZTA components here to provide a theoretical baseline. In subsequent sections, we will adapt and uniquely extend these concepts to address the specific scurity complexities of the O-RAN ecosystem.

\subsection{\MakeUppercase{The NIST 800-207 Special Publication}}

The National Institute of Standards and Technology (NIST) released Special Publication 800-207, establishing the foundational concepts, driving motivations, and core components of ZTA~\cite{stafford2020zero}. This framework is anchored by several defining tenets~\cite{katsis2025zero}:

    \noindent\textbf{Secure communications irrespective of network location.} All network traffic requires robust encryption and authentication, regardless of whether the communication originates internally or externally. Maintaining data integrity and confidentiality is mandatory across all network boundaries.

    \noindent\textbf{Session-based access with least privilege.} System access must be strictly compartmentalized per session. Under the least privilege paradigm, each transaction is granted only the absolute minimum permissions required to complete the specified operation.

    \noindent\textbf{Dynamic access decisions based on multiple factors.} Resource authorization is an active, ongoing evaluation factoring in the requester's identity, the endpoint's current state (e.g., software patch levels), and contextual variables. Contextual elements include behavioral metrics like abnormal access times or unusual data requests, as well as environmental threat indicators and network origin.

    \noindent\textbf{Continuous monitoring of device integrity and security posture.} Enterprises must perpetually audit the health and security posture of all connected endpoints. This mandates deploying systems capable of monitoring asset states and automatically applying patches. Furthermore, policies must distinguish between organizationally managed devices and unmanaged (BYOD or external) hardware, applying significantly more stringent controls or outright denial to the resources.

    \noindent\textbf{Data collection for security enhancement.} Security postures must be continually refined by aggregating and scrutinizing network telemetry. 
    Organizations analyze traffic flows and endpoint activity to build behavioral baselines, allowing them to instantly flag anomalies like unauthorized lateral movement or data exfiltration.

\begin{figure}[htbp]
\centerline{\includegraphics[width=.9\columnwidth]{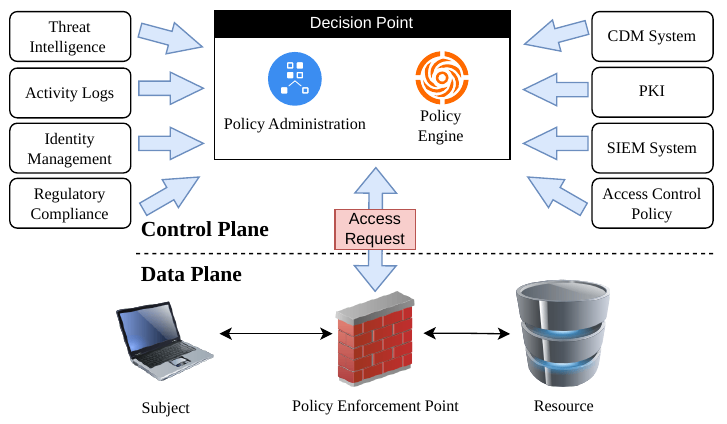}}
\caption{\label{fig:nist-zt-architecture} Logical components of ZTA~\cite{stafford2020zero}. Figure adapted from~\cite{katsis2025zero}.
}
\end{figure}

Figure~\ref{fig:nist-zt-architecture} outlines the fundamental ZTA components detailed in the NIST publication, dividing the infrastructure into two distinct operational layers: the control plane and the data plane. The data plane represents the physical and logical network of interconnected assets, spanning from enterprise laptops and operational technology (OT) sensors to cloud-hosted databases. Conversely, the control plane houses the administrative infrastructure responsible for managing access rules, monitoring the data plane, and enforcing security policies.

Whenever a subject requests resource access, the interaction is intercepted by a PEP (e.g., a firewall system). The PEP relays this request to a decision module, where it is initially evaluated by the PA before reaching the PE. The PE makes the final authorization ruling by synthesizing inputs from multiple intelligence sources~\cite{stafford2020zero,katsis2025zero}:

    \noindent\textbf{Access Control Policy.} 
    The primary baseline for authorization, consisting of manually or automatically generated rules that define the scope of permitted actions across the network and individual end-systems.
    
    \noindent\textbf{Activity Logs.} 
    Detailed records capturing both network-level telemetry (e.g., traffic flows) and end-system interactions (e.g., application usage). The PE analyzes these logs to identify anomalous patterns, while the PA can use real-time log ingestion to sever active connections if malicious behavior emerges.

    \noindent\textbf{Threat Intelligence.} 
    External or internal feeds detailing emerging vulnerabilities and active attack campaigns. This intelligence allows the PE to dynamically override existing policies, proactively denying or revoking access if a new threat compromises a previously trusted entity.
    
    \noindent\textbf{Continuous Diagnostics and Mitigation (CDM) System.} 
    A component dedicated to tracking the real-time software and configuration state of network assets. If an endpoint is running deprecated or vulnerable software, the PE can block access until the CDM system successfully applies the necessary updates.

    \noindent\textbf{Regulatory Compliance.} 
    Directives ensuring the infrastructure adheres to legal frameworks like the GDPR~\cite{gdpr}, FISMA~\cite{fisma}, or CCPA~\cite{ccpa}. These rules might mandate automated data anonymization or enforce strict retention limits to protect sensitive subject information.
    
    \noindent\textbf{Public Key Infrastructure (PKI).} 
    The cryptographic backbone responsible for minting, distributing, and overseeing the digital certificates used to authenticate services, applications, and hardware devices during secure exchanges. This infrastructure can be hosted internally or managed by a trusted third party.

    \noindent\textbf{Identity Management System (IMS).} 
    As detailed in Section~\ref{sec:ID-management}, this system acts as the central repository for user profiles, roles, and identity artifacts. It often integrates with the PKI and can operate within a federated framework to seamlessly authenticate external partners~\cite{Le23}.
    
    \noindent\textbf{Security Information and Event Management (SIEM) system.} 
    The central aggregation hub for enterprise-wide security logs. By correlating data from across the IT landscape, SIEMs provide real-time threat detection, generate actionable alerts, and accelerate incident response procedures to mitigate vulnerabilities before they impact critical infrastructure.

The PE processes telemetry from these diverse systems through a trust algorithm to calculate a dynamic risk score for access governance. The underlying evaluation module may employ deterministic logic or utilize modern artificial intelligence and machine learning models to drive decisions based on training data~\cite{Wang24, katsis2024zt}.

\subsection{\MakeUppercase{The Zero-Trust Maturity Model by CISA}}

Building upon the theoretical foundation established by NIST, the Cybersecurity and Infrastructure Security Agency (CISA) published the Zero Trust Maturity Model (ZTMM) to offer a practical deployment roadmap for organizations~\cite{CISA2023ZTMM}. This framework categorizes operations into five foundational pillars: identity, devices, networks, applications and workloads, and data. These are unified by three cross-cutting operational capabilities: governance, automation and orchestration, and visibility and analytics.

The ZTMM describes a four-stage maturity journey, enabling organizations to advance incrementally from static, manual security toward dynamic, automated defenses. At the traditional stage, security relies on siloed controls, static configurations, and limited visibility. The initial stage introduces partial automation, early integration across pillars, and aggregated visibility. In the advanced stage, controls become more automated and coordinated, with centralized monitoring and enforcement. Finally, the optimal stage represents a fully automated posture in which just-in-time and least-privilege access, dynamic policies, and cross-pillar interoperability support comprehensive situational awareness.

To facilitate gradual modernization, the ZTMM defines a four-phase maturity spectrum, guiding enterprises away from rigid, manual operations toward highly adaptive, automated security postures. In the \textit{traditional} phase, architectures rely heavily on fragmented visibility, manual configurations, and isolated security boundaries. The \textit{initial} tier introduces foundational automation, basic cross-pillar integration, and unified logging. Progressing to the \textit{advanced} stage brings centralized policy enforcement and sophisticated, coordinated security controls. Finally, the \textit{optimal} state achieves a fully frictionless, automated environment characterized by dynamic policy rendering, continuous monitoring, just-in-time access, and seamless interoperability across all pillars.

Within this paradigm, specific technical capabilities mature significantly across each pillar. For example, identity verification shifts from static passwords to robust, phishing-resistant multi-factor authentication, eventually culminating in continuous risk evaluation. Endpoint management transitions from basic manual inventory tracking to automated, risk-aware supply chain analytics. Network defense moves past static perimeter firewalls into intelligent micro-segmentation and universally encrypted traffic flows. Applications transition from isolated on-premises deployments to internet-facing services deeply integrated with automated CI/CD security pipelines. Finally, data protection evolves from static access lists to dynamic, attribute-driven classification and real-time exfiltration prevention mechanisms.

%% file: sections/oran.tex
\section{\MakeUppercase{The 5G Network Fundamentals}}
\label{sec:5g_foundations}

\begin{figure}[htbp]
\centerline{\includegraphics[width=\columnwidth]{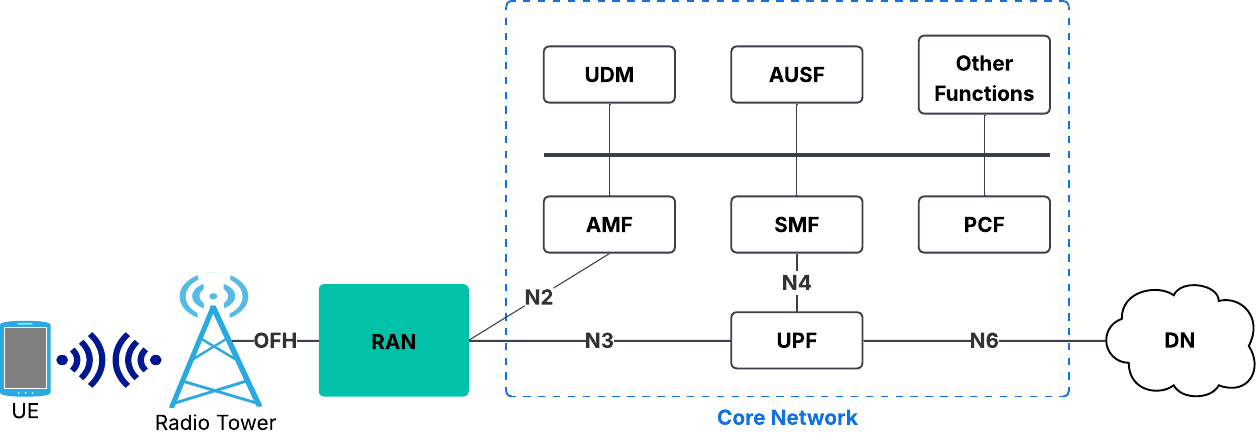 }}
\caption{\label{fig:5g-core} An overview of the 5G network topology.
}
\end{figure}

This section provides foundational background on mobile networks, with emphasis on 5G as the prevailing operational generation. We first outline how user equipment (UE) attaches to the network and how security policies are enforced in the 5G core. We then describe the O-RAN architecture through which UE traffic is propagated from the radio access network (RAN) to the core (Figure\ref{fig:5g-core}). Finally, we present the deployment scenarios that motivate the discussion and research directions in this paper.

\noindent\textbf{Subscriber Identity Authentication.} Identity authentication is a foundational security procedure designed to verify the legitimacy of a UE before allowing access to network services. The authentication framework is defined by 3GPP in the TS 33.501~\cite{3gpp.33.501}.
The primary goal of authentication is to establish a secure and trusted identity for the UE using the Subscription Permanent Identifier (SUPI), a globally unique and permanent identifier (known as IMSI in earlier generations). 

The 5G authentication procedure is coordinated by the Authentication Server Function (AUSF), in collaboration with the AMF and the UDM.
AUSF runs the Authentication and Key Agreement (AKA) protocols namely, 5G-AKA or EAP-AKA, depending on operator configuration to mutually authenticate the UE and the network and derive a shared cryptographic key hierarchy.
These keys are subsequently used by the AMF and the RAN to secure Non-Access Stratum (NAS) and access stratum signaling.
Hence, existing mechanisms primarily verify the subscriber’s legitimacy rather than the trustworthiness of the device hardware or user identity.

\noindent\textbf{Session Management.} Following a successful subscriber authentication session, management is initiated to establish a data connection between the UE and a target Data Network (DN). This process is centered around the establishment, modification, and release of PDU (Protocol Data Unit) sessions, which carry the actual user data traffic in 5G.

Upon authentication, the AMF finalizes the UE registration process and facilitates the signaling required to initiate a PDU session. The UE issues a PDU Session Establishment Request containing parameters such as the requested session type (e.g., IPv4, IPv6, Ethernet), the Data Network Name (DNN), and the Single Network Slice Selection Assistance Information (S-NSSAI). This request is routed to the Session Management Function (SMF) via the AMF.

The SMF serves as the anchor point for session control and orchestrates a variety of critical tasks, including IP address allocation, QoS provisioning, policy enforcement, and the setup of User Plane Function (UPF) forwarding rules. During this process, the SMF interacts with the Policy Control Function (PCF) to retrieve subscriber-specific policies.

\noindent\textbf{Policy Deployment in the Core.} The PCF is responsible for making dynamic policy decisions based on the user's subscription profile, which it obtains from the UDM, as well as session context provided by the SMF. These policies may include constraints on bandwidth, service accessibility, QoS parameters, charging models, and roaming conditions. For example, an enterprise subscriber may be granted access to a high-priority, low-latency network slice with guaranteed bit rates, while a consumer user may be restricted to best-effort traffic handling. The PCF returns the relevant rules to the SMF, which translates them into actionable configurations at the UPF, thereby enforcing policy decisions on the user plane.

Once the session parameters are finalized, the AMF coordinates with the RAN (e.g., gNodeB) to establish the corresponding radio bearers and configure the NG-U (N3 interface) tunnel between the gNB and UPF. Upon successful setup, the session is activated, and the UE is able to transmit and receive data through the established PDU session.

\section{\MakeUppercase{Open Radio Access Network}}
\label{sec:oran}

3GPP provides the foundational specifications for 5G New Radio (NR), including the architecture of the gNodeB and the functional split options between RAN components.
In particular, the 3GPP Technical Report 38.801~\cite{3gpp-tr-38.801} outlines several functional split options (e.g., Option 2, Option 7.2x) and standardizes interfaces such as F1, NG, and Xn for communication between and within gNodeB components. However, the disaggregated architecture comprising the O-RAN Central Unit (O-CU), Distributed Unit (O-DU), and Radio Unit (O-RU), along with the introduction of the Near-Real-Time and Non-Real-Time RAN Intelligent Controllers (Near-RT RIC and Non-RT RIC), is defined solely by the O-RAN Alliance~\cite{oran_alliance, oran-architecture-v13}. The O-RAN Alliance extends the 3GPP architecture by specifying open and interoperable interfaces such as E2, A1, O1, and O2, which facilitate intelligent control, virtualization, and vendor interoperability across the RAN. These specifications are publicly available through the O-RAN Alliance and aim to foster a modular, cloud-native, and AI/ML-enabled RAN ecosystem.

\noindent\textbf{The O-RAN Architecture.} Figure~\ref{fig:oran-architecture} shows the different components of the O-RAN architecture. The gNodeB (gNB) is split into the following components:

\begin{figure}[htbp]
\centerline{\includegraphics[width=.8\columnwidth]{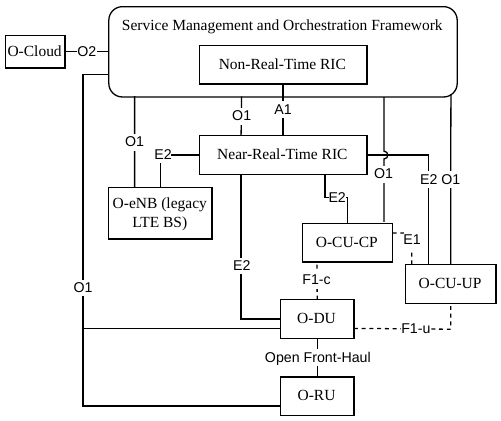}}
\caption{\label{fig:oran-architecture} The O-RAN architecture~\cite{polese2023understanding, oran-architecture-v13}. The solid lines denote the communication interfaces specified by the O-RAN Alliance. Dotted lines represent interfaces specified by 3GPP.
}
\end{figure}

\noindent$\bullet$ \textit{\underline{Control Unit (CU):}}
The CU manages Radio Resource Control (RRC) functions, including connection initiation/termination, QoS, encryption, and error handling. It is decoupled into the Control Plane (CU-CP) and User Plane (CU-UP). The CU-CP handles RRC and relays UE-AMF signaling via the NG-C interface. The CU-UP executes the Service Data Adaptation Protocol (SDAP) and Packet Data Convergence Protocol (PDCP), connecting to the UPF via the NG-U interface. In O-RAN, the interoperable O-CU-CP and O-CU-UP enable vendor-neutral architectures. Crucially, O-CUs act as E2 nodes, connecting to the Near-RT RIC via the E2 interface to stream telemetry and execute xApp-driven control actions (e.g., load balancing, mobility, and slice management). Furthermore, they participate in hierarchical, closed-loop optimization by receiving policies and ML models from the Non-Real-Time RIC indirectly through the Near-RT RIC. CUs can be deployed on general-purpose servers at the network edge or cloud.

\noindent$\bullet$ \textit{\underline{Distributed Unit (DU):}}
The DU handles latency-sensitive, real-time radio functions. It connects to the CU via the F1 interface and the Radio Unit (RU) via the Open Fronthaul interface (typically a 7.2x split). By executing lower protocol layers (e.g., Radio Link Control (RLC), Medium Access Control (MAC), and upper Physical (PHY) layers like modulation, scrambling, and channel coding~\cite{3gpp-ts-38.401}), the DU supports time-critical scheduling, hybrid automatic repeat request (HARQ) processing, and resource allocation. The DU operates in close synchronization with the RU, which manages lower PHY and RF functions (Fast Fourier Transform, beamforming, digital-to-analog conversion). To meet stringent 5G Ultra-Reliable Low Latency Communications latency demands, DUs are deployed on general-purpose servers at the network edge, far-edge, regional clouds, or on-premises. In O-RAN, the interoperable O-DU supports intelligent control by connecting to the Near-RT RIC via the E2 interface.

\noindent$\bullet$ \textit{\underline{Radio Unit (RU):}}
Residing at the network periphery, the RU interfaces directly with the antenna system. Formalized by the O-RAN Alliance, it manages lower PHY layer and Radio Frequency (RF) processing. The RU executes critical time-domain operations, including Digital Front-End (DFE) processing, Fast Fourier Transform (FFT), cyclic prefix manipulation, precoding, beamforming, and digital-to-analog conversion. It connects to the DU via the O-RAN Open Fronthaul interface (typically utilizing the 3GPP Option 7.2x functional split). By standardizing these interfaces, the O-RU achieves multi-vendor interoperability with O-DUs. Typically implemented on FPGAs or programmable ASICs and deployed at outdoor cell sites or rooftops, the RU's separation from the DU enables centralized baseband processing and optimizes RAN deployment for cost, performance, and coverage.

The O-RAN architecture provides two controllers: 
    
\noindent$\bullet$ \textit{\uline{Near Real-Time Radio Intelligent Controller (Near-RT RIC):}}
It supports control loops operating at sub-10 millisecond timescales, including those approaching the microsecond regime. While the current O-RAN specifications only formally define the Non-Real-Time RIC and Near-Real-Time RIC, also the RT-RIC has been proposed in ongoing research and by task forces such as the O-RAN Alliance's Next Generation Research Group (nGRG) as a forward-looking component designed to address ultra-low-latency use cases anticipated in future 5G Advanced and 6G networks~\cite{oran_ngrg_ric_2025}. The RT-RIC is envisioned to reside within or adjacent to the RAN nodes such as O-DUs or even O-RUs and to directly manage time-critical functions such as beamforming, symbol-level scheduling, fast link adaptation, and physical layer parameter optimization. Unlike the Near-RT RIC, which communicates with RAN nodes via the E2 interface, the RT-RIC would require lightweight, high-speed control interfaces capable of operating within the real-time constraints of the physical layer. Given the extreme latency sensitivity of its functions, the RT-RIC must be tightly integrated with hardware accelerators and radio front-end components, potentially leveraging FPGA- or ASIC-based execution environments. 
\noindent$\bullet$ \textit{\uline{Near Real-Time Radio Intelligent Controller (Near-RT RIC):}} 
It is a standardized controller responsible for the near-real-time control and optimization of RAN elements via fine-grained data collection and actions over the E2 interface, supporting control loops operating at timescales between 10 milliseconds and 1 second. The Near-RT RIC hosts third-party applications (xApps) that govern functions such as radio resource management, mobility optimization, and network slicing. While current O-RAN specifications only formally define the Non-Real-Time RIC and this Near-RT RIC, a separate Real-Time RIC (RT-RIC) has been proposed in ongoing research by task forces such as the O-RAN Alliance's Next Generation Research Group (nGRG) as a forward-looking component designed to address ultra-low-latency use cases anticipated in future 5G Advanced and 6G networks~\cite{oran_ngrg_ric_2025}. This proposed RT-RIC would support control loops operating at sub-10 millisecond timescales, including those approaching the microsecond regime. The RT-RIC is envisioned to reside within or adjacent to the RAN nodes such as O-DUs or even O-RUs to directly manage time-critical functions such as beamforming, symbol-level scheduling, fast link adaptation, and physical layer parameter optimization. Unlike the Near-RT RIC, the RT-RIC requires lightweight, high-speed control interfaces capable of operating within the real-time constraints of the physical layer. Given the extreme latency sensitivity of its functions, the RT-RIC must be tightly integrated with hardware accelerators and radio front-end components, potentially leveraging FPGA- or ASIC-based execution environments.

\noindent$\bullet$ \textit{\uline{Non-Real-Time Radio Intelligent Controller (Non-RT RIC):}}  
It is a foundational component of the O-RAN architecture, responsible for enabling centralized intelligence, policy-based guidance, and life-cycle management of machine learning (ML) models across the RAN. Operating on control loops with timescales exceeding one second, the Non-RT RIC is hosted within the Service Management and Orchestration (SMO) framework, typically in a centralized or regional cloud environment. It facilitates RAN optimization through the execution of radio applications known as rApps, which perform functions such as network slicing orchestration, SLA assurance, traffic steering policy generation, and long-term analytics. To support these applications, the Non-RT RIC exposes the R1 interface, which provides a standardized API boundary between the RIC framework and rApps. Through R1, rApps can consume framework services such as data exposure, analytics, and policy management, or register their own capabilities for discovery by other rApps~\cite{oranR1GAP2025}. R1 not only enables secure interaction between the RIC framework and rApps, but also serves as a mechanism for inter-rApp communication mediated by the framework, thereby supporting multi-vendor interoperability and coordinated optimization.  

The Non-RT RIC communicates with the Near-RT RIC via the standardized A1 interface, which it uses to transmit policies, ML model updates, and optimization objectives. Additionally, it interfaces with network functions through the O1 interface for configuration, fault, and performance management, and with the cloud infrastructure via the O2 interface to support the orchestration and monitoring of the O-Cloud environment. By decoupling high-level control from time-sensitive radio operations, the Non-RT RIC enables scalable, vendor-neutral RAN automation, enhances operational efficiency, and provides a platform for integrating AI/ML-driven intelligence into the mobile network. Its role is crucial in supporting hierarchical, closed-loop control loops in collaboration with the Near-RT RIC, thereby realizing the vision of a fully programmable and autonomous RAN.

Last but not least, the O-RAN Cloud infrastructure (O-Cloud, for short) is a foundational element in the O-RAN architecture that provides the virtualized computing, storage, and networking resources required to host all O-RAN software functions, including the O-CU, O-DU, Near-RT RIC, and elements of the SMO framework. The O-Cloud is not a standalone network function but rather an abstraction layer over a cloud-native infrastructure platform designed to support telco-grade workloads with deterministic performance, multi-tenancy, and hardware acceleration capabilities. Its primary role is to enable flexible deployment of RAN components across centralized, regional, and edge clouds, allowing operators to optimize for cost, latency, and energy efficiency.

In the O-RAN system, the O-Cloud interacts with the SMO via the O2 interface, which enables resource discovery, infrastructure orchestration, and lifecycle management of cloud-native network functions. The SMO leverages this interface to provision virtual machines or containers, monitor the health and performance of cloud resources, and enforce infrastructure-level policies such as placement constraints or affinity rules. The O-Cloud must adhere to strict performance and reliability constraints to support latency-sensitive RAN functions, particularly when hosting Near-RT RICs or O-DUs at the edge. It also supports integration with hardware accelerators (e.g., FPGAs, GPUs, SmartNICs) via standardized APIs to offload intensive functions like encoding/decoding or beamforming.

\subsection{\MakeUppercase{Handover Procedures in O-RAN and 5G Architectures}}

\begin{figure}[htbp]
\centerline{\includegraphics[width=0.4\columnwidth]{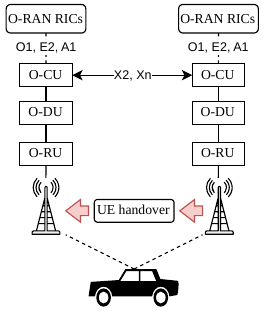 }}
\caption{\label{fig:handover} UE Handovers between RANs.
}
\end{figure}

In O-RAN and 5G NR system deployments, multiple CU instances often coexist to manage geographically or logically distinct subsets of the RAN. Each CU instance independently controls one or more DUs and their associated RUs through the F1 interface. When a UE moves from a cell under the management of one CU instance to a cell managed by a different CU, the system must perform an \textit{inter-CU handover} (see Figure~\ref{fig:handover}).

An inter-CU handover requires explicit coordination between the source and target CUs. To enable this, the 3GPP specifications define the Xn interface~\cite{3gpp-ts-38.423}, which facilitates both control and user plane communication between different gNBs, including their CU components. The Xn interface supports fast handover procedures by allowing the transfer of UE context and signaling information without direct involvement of the core network, thus reducing handover latency compared to core-assisted procedures.

Of course, if the two RANs are managed by the same CU instance, an \textit{intra-CU handover} occurs. In this scenario, the handover procedure can be handled entirely within the CU, without the need for external signaling over the Xn interface. According to 3GPP TS 38.401~\cite{3gpp-ts-38.401}, the CU is responsible for maintaining the UE context during mobility procedures across its managed DUs by utilizing the F1 Application Protocol (F1-AP)~\cite{3gpp-ts-38.473}.

\subsection{\MakeUppercase{RAN-to-Core Communication in O-RAN and 5G Architectures}}

\begin{figure}[htbp]
\centerline{\includegraphics[width=.9\columnwidth]{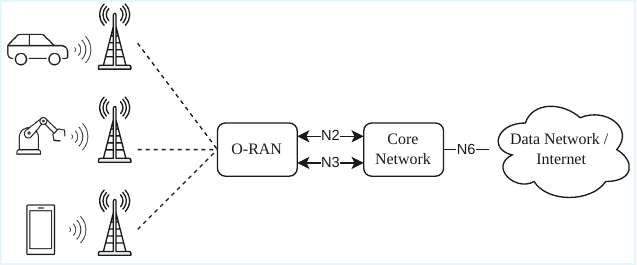 }}
\caption{\label{fig:ran-to-cn} O-RAN to CN communication.
}
\end{figure}

The communication between the RAN and the Core Network is defined by two logical interfaces: N2 and N3~\cite{3gpp-ts-23.501} (see Figure~\ref{fig:ran-to-cn}). From the RAN perspective, these interfaces are referred to as NG-C (control plane) and NG-U (user plane), respectively~\cite{3gpp-ts-38.410}. Specifically, the N2 interface connects the CU-CP to the AMF and carries control signaling related to UE registration, session management, and mobility procedures. The N3 interface connects the CU-UP to the UPF and is responsible for transporting user data traffic using GTP-U tunnels. The UPF then routes it to the Internet or to the appropriate data network.

\section{\MakeUppercase{O-RAN Deployment Scenarios}}
\label{sec:oran-deployment-scenarios}

O-RAN architectures allow a more flexible and disaggregated deployment of cellular networks. While many mobile network operators (MNOs) are adopting O-RAN within their own networks, including greenfield deployments (e.g. Dish’s new 5G network) and brownfield upgrades of existing sites~\cite{5GAm_Trends_OpenRAN_2024}, O-RAN also enables novel deployment scenarios involving multiple stakeholders. Below we outline the key O-RAN deployment scenarios.

\subsection{\MakeUppercase{Traditional Operator-Owned O-RAN Deployment}}

In this scenario, a single MNO owns and operates the entire RAN and core network using O-RAN compliant equipment. This is the classic public network model, where the operator deploys O-RAN components (O-RU, O-DU, O-CU, RIC, etc.) across one or multiple geographic areas for its subscribers. Such deployments can be greenfield (a new network built entirely with O-RAN, as seen with Rakuten~\cite{RakutenSymphony_OpenRAN} or Dish~\cite{DISH_OpenRAN2023}) or brownfield (integration of O-RAN units into an existing network). 
This is often described as the ``MNO owned and operated'' model~\cite{ORAN_PrivateNetworks_2025} as one viable category for both private and public network deployments. In practice, this scenario represents an evolution of traditional RAN rollouts, but with open interfaces and multi-vendor components that align with O-RAN specifications.

\subsection{\MakeUppercase{Private Cellular O-RAN Networks}}

Another scenario is a fully private 5G network where one entity (e.g., an enterprise, university campus, or factory) deploys and controls both the O-RAN based RAN and its own core. In this isolated non-public network model, the enterprise owns all 5G system functions  O-RUs, O-DUs/O-CUs, and a 5G core deployed on its premises~\cite{ORAN_PrivateNetworks_2025}. This corresponds to a Stand-alone Non-Public Network (SNPN) in 3GPP terms. By using O-RAN architecture, a private network gains the benefits of interoperability and flexibility (e.g. mixing vendors for radios and baseband) while tailoring the network to the enterprise’s specific requirements (low latency control, local breakout, security, etc.). This scenario is essentially a self-contained O-RAN deployment serving a single organization’s needs, with no dependency on public mobile operators.

\subsection{\MakeUppercase{Hybrid Public-Private Deployment (Separate RAN and Core Entities)}}

O-RAN also supports hybrid deployments where separate entities operate the RAN and the core. For instance, an enterprise or neutral-host (e.g., a university campus) can deploy an on-site O-RAN that connects to a Mobile Network Operator’s (MNO) 5G core via the standard 3GPP NG interface. Often categorized as a Public Network Integrated NPN (PNI-NPN), this model links a local enterprise RAN to a public PLMN. In this arrangement, the enterprise acts as a third-party access network in its own administrative domain, managing local radio resources without needing to operate a proprietary core. Meanwhile, the MNO’s core handles user authentication, mobility management, and subscriber services. This model offers two primary advantages: first, it efficiently improves coverage and capacity in specific venues (campuses, buildings, rural areas) using O-RAN’s openness; second, it allows the local O-RAN operator to define custom security policies (e.g., access control), QoS policies, and localized monitoring systems like anomaly detectors.

\subsection{\MakeUppercase{Multi-Operator RAN Sharing with O-RAN}}

\begin{figure}[htbp]
\centerline{\includegraphics[width=.8\columnwidth]{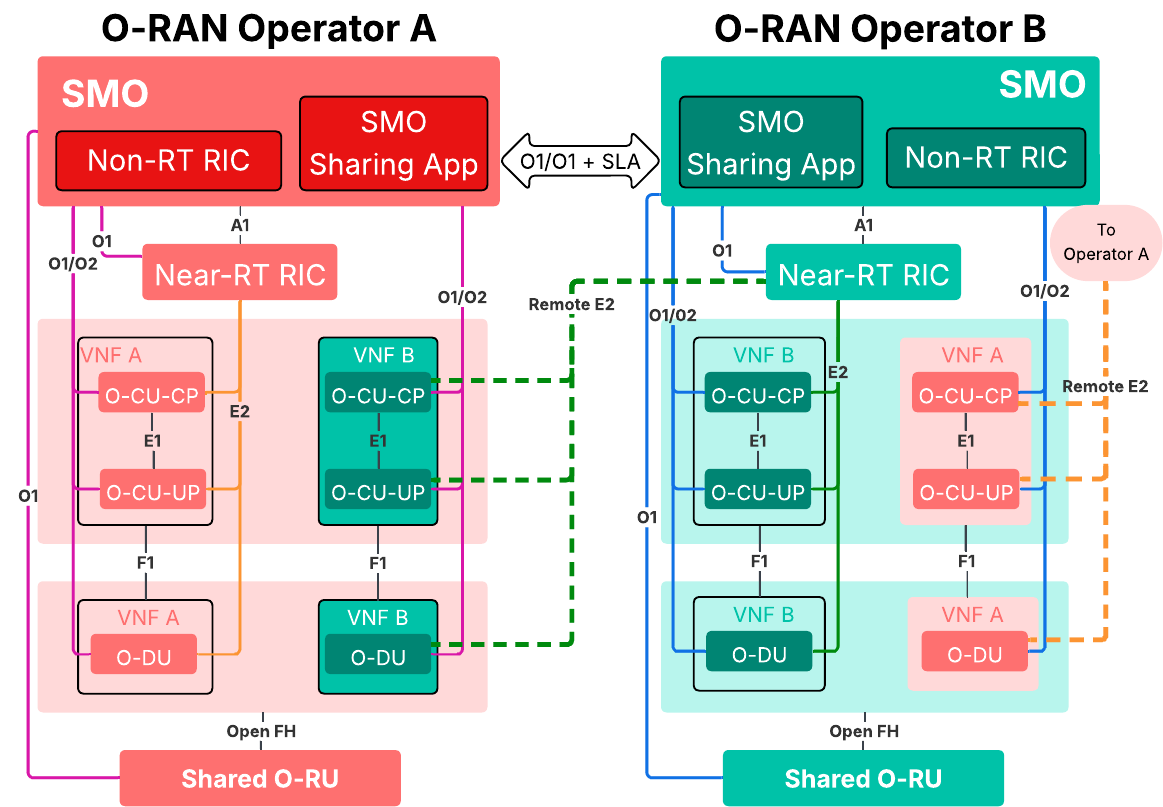 }}
\caption{\label{fig:oran-resource-sharing} Resource Sharing in O-RAN~\cite{oran2025usecases}.
}
\end{figure}

Beyond one-to-one partnerships, O-RAN enables complex multi-operator RAN sharing where a single infrastructure concurrently serves multiple operators or public/private networks. For instance, Operator A can host Operator B’s O-DU/O-CU software as Virtual Network Functions (VNFs) on shared hardware~\cite{oran_wg11_threatmodeling_2025, oran2025usecases, ORAN_WG11_SharedORU_TRR004_v0600_2024}. Both utilize the same radio equipment while maintaining separate core networks. Traditional 3GPP RAN sharing arrangements (MOCN/MORAN) allow operators to share base stations, but they typically require tight coordination and limit each operator’s flexibility. By contrast, the O-RAN model introduces open interfaces and virtualization that greatly ease multi-operator sharing. Standardized interfaces (e.g., Open Fronthaul and E2) allow multiple independent O-DUs or O-CUs to connect to a common O-RU in a neutral-host configuration. Crucially, each operator retains independent control; via the O1, O2, and E2 interfaces, they can continuously monitor performance and optimize shared radio parameters according to their own policies. This granular flexibility makes O-RAN ideal for spectrum sharing and neutral host deployments (e.g., in-building or stadium coverage), far surpassing the capabilities of legacy RAN systems.

Figure~\ref{fig:oran-resource-sharing} illustrates a RAN sharing scenario in which Operator A (the host) provides infrastructure resources that also host the virtualized RAN functions (O-DU, O-CU) of Operator B (the guest). Both operators maintain their own independent SMO frameworks, which include the Non-RT RIC and Near-RT RIC.

Operator A manages its own VNFs directly through local O1/O2 interfaces, as in a conventional O-RAN deployment. Operator B, however, cannot directly access its VNFs at the Operator A's site. Instead, it issues configuration and orchestration requests over the ``O1/O2 remote + SLA'' channel, which is mediated by SMO-sharing Apps deployed on both sides. These applications enforce service-level agreements (SLAs) and filter management operations before they are executed against the hosted VNFs. For near-real-time control, each operator’s Near-RT RIC remains independent and connects to its own VNFs using the E2 interface. Operator B’s Near-RT RIC reaches its VNFs at Opearator A's site via an ``Remote E2'' link, while Operator A uses a direct local E2 connection. In this way, radio optimization, policy enforcement, and performance monitoring remain under the control of each operator.

Overall, the figure highlights the dual management paths in shared RAN deployments: (i) direct O1/O2/E2 control for the host operator, and (ii) remote O1/O2/E2 access with SLA enforcement for the guest operator. This arrangement enables multi-operator RAN sharing while ensuring administrative separation and compliance with negotiated SLAs.

%% file: sections/control_plane_zta.tex
\section{\MakeUppercase{Application of ZTA on the Cellular Control Plane}}
\label{sec:cp-zta}

This section explores key use cases and requirements for applying ZTA to the cellular control plane. These use cases span the entire O-RAN software stack from the procurement of applications and ML models to the enforcement of access controls and monitoring of permissions for the underlying RAN infrastructure.

\subsection{\MakeUppercase{Secure Procurement and Integration of Third-Party Components in the O-RAN Ecosystem}}

The O-RAN ecosystem is designed to support the procurement and integration of external components (often referred to as ``plug-and-play'' modules) that may be developed by third-party vendors outside the control of the O-RAN operator. These components include: (1) xApps (deployed in the Near-RT RIC), (2) rApps (deployed in the Non-RT RIC), (3) dApps (deployed within the RAN itself), and (4) AI/ML models (typically integrated within the Non-RT RIC).

While these components enable rapid innovation and flexibility, they also introduce significant risks. Malicious or compromised components could disrupt RAN operations, leak sensitive data, or exfiltrate information such as user identities and communication patterns. As such, securing the procurement and integration process for these components is critical to safeguarding the integrity and reliability of the O-RAN ecosystem. This requirement is well aligned with the strategic objectives of ZTA as outlined in guidelines such as those from the U.S. Department of Defense~\cite{dodZT} and the ZTMM by CISA~\cite{CISA2023ZTMM}.

\textbf{Secure Software Procurement.} 
A secure software procurement process involves several essential verification steps~\cite{rais2024zero}:

\begin{itemize}[leftmargin=*, itemsep=0.1em]
    \item \textit{Source Code Verification}: Ensuring that the software source code, whether manually written or generated by LLMs, is secure and trustworthy.
    
    \item \textit{Software Build Integrity Verification}: Confirming that the software build process is reproducible and resilient to supply chain attacks by testing against diverse inputs and verifying the consistency of outputs.

    \item \textit{Secure Software Distribution}: Safeguarding the software distribution pipeline to prevent unauthorized alterations of artifacts intended for deployment.

    \item \textit{Artifact Verification}: Enabling downstream consumers to verify the integrity and authenticity of distributed binaries through cryptographic methods such as hashing and digital signatures. This ensures that software artifacts are not tampered with after release and allows users to validate the publisher's identity by verifying the signature against their public key.
\end{itemize}

In the context of O-RAN, the security aspects of software procurement and integration remain underdeveloped. Current O-RAN technical specifications do not provide detailed procedures for secure software procurement. For example, the O-RAN Near-RT RIC APIs specification~\cite{oran2024ricapi} outlines the xApp registration process (Section 8.2.1); however, it does not define any security mechanisms to support the secure procurement steps discussed earlier. Similarly, the O-RAN Near-RT RIC Architecture specification~\cite{oran2025ricarch} mentions that the Near-RT RIC should authenticate xApps, but it does not provide specific guidance or technical details on how such authentication should be performed.

Recognizing this gap, the O-RAN Security Working Group (WG11) has identified the secure onboarding of xApps as a critical challenge and has proposed an extended solution to enhance the security of the xApp registration process~\cite{oran2024security}.

\begin{figure}[htbp]
\centerline{\includegraphics[width=.7\columnwidth]{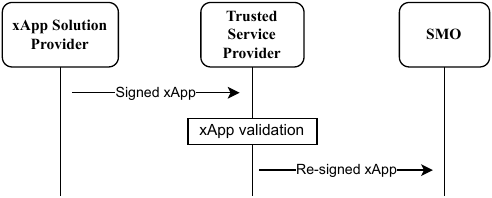}}
\caption{\label{fig:secure-xapp-onboarding} Secure xApp onboarding process via SMO~\cite{oran2024security}.}
\end{figure}

In this proposed solution, the Service Management and Orchestration (SMO) framework plays a central role in the secure onboarding and deployment of xApps. During the onboarding phase, the xApp Solution Provider (the entity responsible for developing the xApp) submits the xApp package to a Trusted Service Provider, a publicly trusted entity that has an established trust relationship with the SMO. The Trusted Service Provider performs several critical checks, including the digital signature verification for package integrity, the operational testing to ensure correct behavior, and schema validation for compliance with expected formats.

Upon successful completion of these validations, the SMO signs the xApp package and adds it to the service catalog, making it available for deployment within the O-RAN environment. This process is illustrated in Figure~\ref{fig:secure-xapp-onboarding}.

Once the onboarding phase is complete, the SMO coordinates the deployment of the xApp to the Near-RT RIC. In this stage, the SMO handles resource allocation and orchestrates the deployment process to ensure that the xApp is properly instantiated on the Near-RT RIC platform.


\begin{figure}[htbp]
\centerline{\includegraphics[width=\columnwidth]{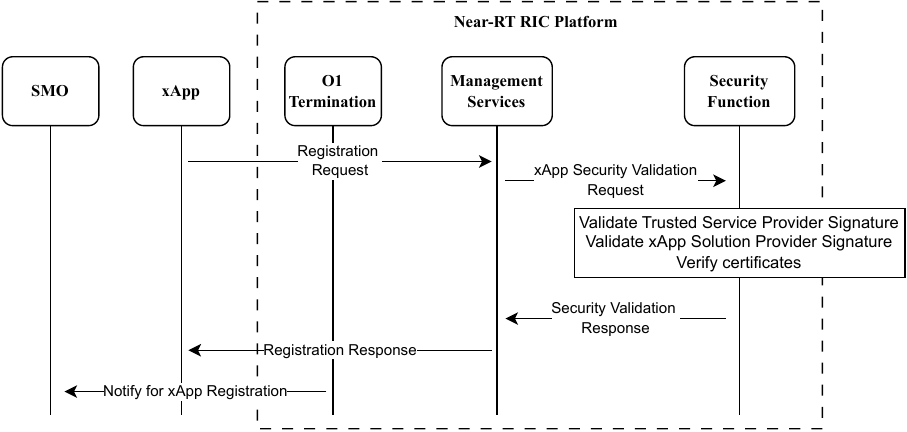}}
\caption{\label{fig:secure-xapp-registration} Extended xApp registration procedure with Near-RT RIC~\cite{oran2024security}.}
\end{figure}

Following deployment, the xApp independently initiates its registration process with the Near-RT RIC management services. This step is entirely managed within the Near-RT RIC.
During registration, the Near-RT RIC performs additional integrity checks by verifying the digital signatures provided by both the Trusted Service Provider and the Solution Provider. It also validates the associated certificates, ensuring that none have been revoked. Only after all security checks have passed does the Near-RT RIC finalize the xApp registration, granting it access to platform APIs and other system components. Upon successful registration, the Near-RT RIC notifies the SMO to complete the process.

While the proposals by the O-RAN Security Working Group represent important steps toward establishing a secure procurement process for xApps, they have several limitations. First, these proposals have not yet been incorporated into official standards, as they remain in the realm of suggested solutions. Second, they are often vague; for example, they do not clearly define how the Trusted Service should perform behavioral checks on xApp software prior to signing. Third, the proposed measures do not comprehensively cover all phases of the procurement process.

It is also critical to understand that verifying the digital signature of an xApp only authenticates the origin of the software by confirming that a trusted entity, such as the xApp provider or a designated signer, has issued the package. However, this signature says nothing about the security or integrity of the software itself.

Therefore, a secure procurement process must go beyond signer authentication. It should incorporate robust verification mechanisms that validate both the software build process and the underlying source code to ensure the software has not been maliciously altered at any stage of the supply chain. A notable example highlighting this need is the SolarWinds supply chain attack, in which attackers injected malicious code prior to the build and signing process~\cite{bertino2023machine}. As a result, the Trusted Service ended up signing a compromised build, inadvertently legitimizing the attack.

In short, verifying signatures and certificates is necessary but not sufficient. Comprehensive security must also include verification of the software’s integrity throughout the entire development and distribution pipeline.

\begin{ZTATakeaways}
  \item The O-RAN ecosystem requires a verifiable chain of trust extending beyond certificate validation encompassing build reproducibility, behavioral verification, and provenance of code to meet ZTA’s ``never trust, always verify'' requirements.

    \item Embedding ZTA principles would shift onboarding from static certification to continuous assurance, where validation persists across development, deployment, and runtime stages.

    \item ZTA exposes a structural gap in current O-RAN standards: trust decisions are centralized in the SMO without continuous verification, leaving post-onboarding drift or compromise unaddressed.

    \item The concept of a ``Trusted Service Provider'' must evolve under ZTA from a one-time signer to a persistent verifier maintaining integrity attestations throughout the xApp’s lifecycle.
\end{ZTATakeaways}

\textbf{Secure rApps/dApps Procurement.}
The O-RAN Security Working Group acknowledges security concerns surrounding the procurement of rApps in its technical specification~\cite{oran2022nonrtricsecurity}. These concerns include the risk of incorporating malicious or vulnerable rApps into the system, potentially leading to attacks such as unauthorized access, backdoors, or remote code execution. However, the report does not propose concrete mechanisms or high-level frameworks for secure procurement and validation of rApps within the O-RAN architecture.

In contrast, no official technical specifications currently exist for dApps at the time of this writing. Instead, the O-RAN alliance has released a research report~\cite{oran2024dapps} that introduces the concept of dApps as lightweight, distributed, programmable applications deployed on O-RAN Distributed Units (O-DUs) and Centralized Units (O-CU-CP and O-CU-UP). These applications are designed to complement xApps and rApps by enabling localized, data-driven functions such as spectrum management, energy efficiency, traffic classification, and real-time scheduling. The report outlines architectural requirements for dApps, including data flow models between dApps and RAN elements, and discusses potential deployment modes either as standalone dApps or as real-time RICs embedded within the RAN.

Despite this progress in architectural vision, the report does not address the security implications or provide guidance on secure procurement and lifecycle validation of dApps. As such, the secure onboarding, verification, and integrity assurance of dApps remain an open challenge requiring further research and standardization.

\begin{ZTATakeaways}
\item The O-RAN Security Working Group acknowledges that rApp procurement introduces security risks, including the possibility of malicious or vulnerable applications leading to unauthorized access, backdoors, or remote code execution.

\item Current specifications do  not include concrete mechanisms or frameworks for secure procurement and validation of rApps.

\item  For dApps, no official technical specifications currently exist. As a result, secure onboarding, verification, and integrity assurance of dApps remain an open challenge, requiring additional research and standardization.
\end{ZTATakeaways}

\textbf{Secure AI/ML Procurement.}
While AI/ML models in O-RAN may be trained within the O-RAN environment itself, it is also possible for these models to be developed externally either by third-party vendors or in other training environments and then procured for direct deployment or fine-tuning within the target O-RAN system.

The O-RAN Security Working Group has released a technical report addressing the security of AI and ML within O-RAN systems~\cite{oran2025aimlsecurity}. The document outlines a range of threats based on the OWASP Top 10 security risks for ML systems, including input manipulation, data poisoning, and model poisoning attacks. Of particular relevance to the model procurement process are AI supply chain attacks, which target vulnerabilities in the lifecycle of AI model development, distribution, and deployment.

These attacks pose a significant threat to the integrity of O-RAN systems. This is due to the fact that many decisions made within the control plane and by extension, actions executed in the data plane are influenced by the output of AI models. For example, consider an xApp designed for intelligent traffic steering that relies on an LSTM model to predict future traffic load based on telemetry data collected from the RAN. Depending on the predicted load, the xApp may trigger a handover decision to redirect a UE to a less congested cell in order to preserve Quality of Service (QoS). If the model is compromised, however, the consequences can be severe. A poisoned or manipulated model could intentionally mispredict traffic, such as forecasting congestion in lightly loaded cells, causing UEs to be handed over to already overloaded cells. This could result in cascading failures and denial of service, as the affected cells become unable to serve the redirected UEs.

The procurement and integration of externally trained AI/ML components into O-RAN thus introduces new attack vectors and expands the system’s threat surface. The O-RAN Working Group 11 highlights several such threats that can compromise the confidentiality, integrity, and availability of AI-driven functions within the architecture, including:

    \noindent \textit{\underline{Data Poisoning:}} Adversaries may manipulate or corrupt training datasets used to develop AI models. By injecting misleading or malicious data, attackers can cause the model to learn incorrect behaviors, make biased predictions, or degrade its overall performance. In some cases, data poisoning can even introduce hidden vulnerabilities or backdoors in the final deployed model.

    \noindent\textit{\underline{Model Tampering:}} During either the training or deployment phase, attackers might tamper with the AI model itself by injecting malicious logic or code. Such tampering can give attackers the ability to control model behavior, steal sensitive inferences or user data, or force incorrect or adversarial outputs on demand.

    \noindent \textit{\underline{Backdoor (Trojan) Attacks:}} In these attacks, a seemingly legitimate AI model is embedded with malicious behavior that remains inactive until a specific trigger is encountered (e.g., a particular input pattern). Once triggered, the backdoor can bypass security mechanisms, leak information, or subvert the intended model behavior.
    These attacks are especially dangerous due to their stealthy nature and ease of remote activation.

    \noindent \textit{\underline{Hardware Tampering:}} Attackers may target the underlying hardware used to train or run AI models such as GPUs, FPGAs, or AI accelerators. By embedding hardware-level backdoors or vulnerabilities, malicious actors can compromise AI systems regardless of the software's integrity, affecting the system's security at a fundamental level.

    \noindent \textit{\underline{Insecure Maintenance APIs:}} AI systems often expose APIs for deployment, maintenance, and monitoring. If these APIs lack proper access control and authentication, they can be abused by attackers to inject malicious updates, retrieve confidential data or manipulate the AI system remotely.
    This creates a significant risk for AI/ML-enabled components in an open and disaggregated architecture like O-RAN.

    \noindent\textit{\underline{Vulnerable Third-Party Libraries and Dependencies:}} AI solutions frequently depend on external open-source libraries and software components. If these dependencies are not adequately vetted or kept up-to-date, they can introduce exploitable vulnerabilities or even deliberate backdoors into the AI system. A compromised or unverified software supply chain significantly increases the system’s attack surface.

    \noindent\textit{\underline{Evasion and Input Manipulation Attacks:}} As highlighted in the O-RAN WG11 Technical Report on AI/ML Security, adversaries can alter input data during the inference phase to mislead the model. In the context of wireless networks, this translates to signal-level adversarial evasion attacks. Attackers can apply small, mathematically optimized perturbations to over-the-air signals, which manifest as manipulated telemetry (e.g., perturbed in-phase and quadrature - I/Q - samples) when processed by the O-RU and forwarded to the Near-RT or Non-RT RIC~\cite{oran2025aimlsecurity}. Because xApps and rApps rely on this telemetry, these crafted adversarial examples can cause significant decision errors, such as tricking a traffic-steering model into misallocating resources. The WG11 report acknowledges that attackers without prior knowledge of the model (black-box attacks) can still successfully craft these adversarial examples by querying the model or using transfer-based evasion techniques.

    Furthermore, recent academic studies demonstrate that dynamic and reshaped attack patterns at the lower layers can successfully deceive ML-based xApps. For example, El Houda et al.~\cite{abou2024federated} note that AI-based systems deployed to mitigate signal-level jamming attacks are inherently vulnerable to dynamic and zero-day attack patterns, as the adversary continuously alters the radio signal interference to evade the model's learned baseline. Similarly, Groen et al.~\cite{groen2024timesafe} demonstrated experimentally that ML models monitoring the open fronthaul can be easily deceived if the attacker ``reshapes'' the temporal pattern of their exploits; in their study, altering the duration and recovery intervals of a timing attack caused Convolutional Neural Network models to completely fail in detecting the malicious traffic, highlighting the fragility of xApp models against adaptive adversaries. Even physical compromises at the edge, such as repeatedly rebooting an unsecured O-RU's power socket, generate anomalous KPM telemetry (e.g., handover storms) that can deceive ML detection systems if they are not explicitly trained on these novel, physical-layer sub-use cases~\cite{mayhoub2024new}. Addressing these physical-layer and dynamic evasion attacks within a ZTA necessitates implementing robust, adversarial-trained ML models (e.g., utilizing Transformers or Federated Deep Reinforcement Learning) and continuously cross-validating telemetry from multiple distributed O-RUs to detect spatial and temporal anomalies in real time.

The technical report does not propose concrete solutions to address the significant challenges in securing AI/ML components. Instead, it outlines a set of high-level security controls that highlight important considerations for mitigating risks but fall short of detailing how these controls should be implemented. For instance, the report recommends code signing and verification using strong cryptographic signatures to ensure model authenticity and prevent unauthorized tampering during distribution. It also suggests adversarial training, exposing the model to perturbed inputs during training to increase its robustness against evasion attacks. Another suggested control is the adoption of explainable AI models, which can improve interpretability and help detect anomalous or adversarial behavior more effectively.

While these controls align with ZTA principles, they are not sufficient on their own. For example, there is no guidance on how to vet the provenance of training data or detect tampering in a model’s internal logic. Unlike traditional software, which can often be statically analyzed for correctness, AI models encode knowledge in statistical parameters that are inherently opaque. Therefore, model verifiability poses a unique challenge.

One possible enhancement to the procurement process could involve requiring that vendors not only sign the model code and learned parameters, but also provide benchmark datasets and metadata describing the training data sources. Upon deployment, the target O-RAN system could use these benchmarks to validate model behavior in a controlled environment. This could be supplemented with red-teaming approaches or curated adversarial inputs to evaluate the model’s robustness and generalization capacity.

Moreover, the procurement process should also account for lifecycle considerations, including runtime monitoring to detect model drift, behavioral deviations, or post-deployment tampering. Techniques such as model watermarking, behavioral attestation, or reproducibility guarantees through secure enclaves may offer further protection. 
The model watermarking, for example, involves embedding a secret, verifiable signature into the model’s behavior during training, for example, configuring the model to produce specific, predictable outputs in response to carefully crafted trigger inputs. These trigger inputs are imperceptible to normal users and do not affect the model’s performance but can later be used by the model provider or procuring O-RAN operator to verify the model’s authenticity and detect unauthorized modification.
Together, these measures would help ensure that AI/ML models in O-RAN systems maintain their integrity and intended functionality throughout their operational lifecycle.

\begin{ZTATakeaways}


    \item The supply chain for AI introduces a new trust dependency graph (training data → model → deployment hardware), meaning ZTA enforcement must span from data pipelines to inference endpoints.

    \item Suggested ZTA controls (like code signing and explainable AI) are insufficient because AI models are inherently opaque. A robust procurement process should require vendors to provide benchmark datasets and training data metadata for validation and red-teaming. Furthermore, runtime monitoring and techniques like model watermarking (embedding verifiable signatures) are needed to detect post-deployment tampering and model drift.

    \item Beyond software vulnerabilities, AI/ML xApps are susceptible to physical-layer evasion attacks where adversaries manipulate over-the-air signals to feed deceptive telemetry into the control plane. ZTA enforcement must therefore include adversarial training and the continuous cross-validation of telemetry across distributed O-RUs to detect signal-level anomalies.
\end{ZTATakeaways}

\textbf{Secure RAN Component Procurement and Resource Sharing.} 
Beyond the procurement of higher-level software in the O-RAN stack, the openness and flexibility of the architecture extend to the procurement of RAN nodes themselves. An operator may, for example, acquire a CU product from a vendor or deploy it as a cloud-hosted instance in a multi-tenant shared environment. This procurement model significantly expands the security considerations in O-RAN, as ZT principles must apply not only to management software but also to procured RAN components.

The O-RAN Alliance has recognized these risks and issued security reports analyzing the challenges of third-party RAN components that communicate via open, and often insecure, interfaces. In particular, Section 7.4.9 of the O-RAN Threat Modeling and Remediation Analysis report~\cite{oran_wg11_threatmodeling_2025} highlights the case of Shared O-RUs in multi-tenant environments. Shared O-RUs allow multiple MNOs to use the same radio infrastructure, but this introduces complex risks spanning confidentiality, integrity, availability, authentication, and authorization.
\begin{itemize}[leftmargin=*, itemsep=0.1em]
    \item \textit{Confidentiality Risks:} Data-at-rest at the SMO or O-DU, as well as data-in-transit between Shared O-RUs, O-DUs, and SMOs, may be exposed to unauthorized actors if confidentiality protections are weak. Tenants with access to protocol stacks could eavesdrop on unprotected control- or user-plane traffic, capturing information from other tenants.

    \item \textit{Availability Risks:} Shared O-RUs are vulnerable to denial-of-service (DoS) vectors such as modification or deletion of Open Fronthaul (O-FH) control-plane messages, clock hijacking on the S-Plane, and volumetric DDoS attacks from O-DUs or SMOs. Even benign misconfigurations (e.g., conflicting parameters, compromised DHCP/DNS during bootstrapping) can cause service outages.

    \item \textit{Configuration and Role Assignment Risks:} Errors in host role assignment or spectrum allocation may grant unauthorized tenants elevated privileges. In multi-tenant environments, weaknesses in the trust chain increase the risk of privilege inheritance or retention beyond intended scope.

    \item \textit{Elevation of Privilege Threats:} Tenants misusing host or ``sudo'' privileges can gain unauthorized access to data or disrupt service. Spoofing of the SMO to manipulate array-carrier configurations at the O-DU or O-RU, or tampering with O1 messages to conceal failures, represent serious integrity threats.

    \item \textit{Untrusted Peering Threats:} One tenant may act as a malicious peer, exploiting shared infrastructure to access another tenant’s resources or disrupt operations. These cross-tenant risks make strong tenant isolation, trusted execution environments, and peer authentication essential.
\end{itemize}

Even outside shared environments and resources, supply chain threats apply. A malicious or compromised vendor could insert backdoors or weak security controls. For instance, a backdoored O-CU could exfiltrate user data or subtly manipulate traffic. A malicious O-RU or O-CU could also send malformed or mis-sequenced packets over the E2 or O-FH interfaces to disrupt service or gain unauthorized access. These threats are not theoretical: malicious insider activity in telecom increased in 2024~\cite{oran_security_update_2025}, and vulnerabilities have been disclosed, such as CVE-2023-40997, a buffer overflow in the O-RAN Software Community’s \verb|ric-plt-lib-rmr| v4.9.0 that allows remote DoS via crafted packets. Another realistic example is a compromised O-RU or an O-FH switch, both critical open components in the fronthaul, could be manipulated to disrupt service or facilitate unauthorized access. Such threats are not merely theoretical; academic research has demonstrated the feasibility of various attacks under different threat assumptions.  For example, 5G-Muffler, proposed by Lin et al.~\cite{lin20255g}, presents a set of covert DoS attacks specifically targeting the O-FH interface within O-RAN 5G networks. These attacks primarily exploit the random access process, specifically the first message (Msg1) sent from the UE to the network, which contains a random access preamble. By disrupting this initial step, 5G-Muffler effectively prevents UEs from connecting to the 5G network, rendering it inaccessible. A crucial design aspect of these attacks is their stealthiness, as they are engineered to be invisible to anomaly detection mechanisms operating above the PHY layer and leave no abnormal traces in the O-DU or the system behind it.

The O-RAN Alliance has responded by embedding ZT principles into its security requirements~\cite{oran_wg11_srcs_2025}. The Security Requirements mandate mutual TLS (mTLS) for control and management interfaces, role-based access control on O1/O2, support for 802.1X and MACsec on fronthaul links, and secure boot and Software Bill of Materials (SBOM) disclosure to mitigate supply-chain risks. Certification and badging programs further support independent validation of vendor compliance.
To ensure secure deployment, operators must adopt a holistic strategy that spans technical hardening, tenant governance, and procurement policy. All O-RAN interfaces should be secured with authentication, encryption, and integrity protection, while multi-tenant environments must enforce strong isolation to prevent tenants from influencing or accessing each other’s resources. The O-RAN Alliance WG11 has already proposed transport-layer protocols, such as IPSEC and mTLS, for securing the open interfaces~\cite{oran2024zta}.

Continuous monitoring of RAN components is equally critical, including anomaly detection on inter-component messages and systematic log analysis. Importantly, related work in cellular security has already demonstrated the practicality of such anomaly detection: for example, Mubasshir et al.~\cite{mubasshir2024fbsdetector} proposed a system for detecting fake base stations in cellular networks, which illustrates how abnormal signaling can be identified in real time. Similar approaches can be adapted for O-RAN to flag malicious or misbehaving procured components. Finally, procurement processes themselves must embed security requirements by mandating SBOMs, security certifications such as NESAS (by GSMA and 3GPP) or O-RAN compliance badges, timely patch management, and clear vendor accountability. 

\begin{ZTATakeaways}

    \item Shared O-RUs in multi-tenant environments introduce complex risks across confidentiality (eavesdropping), availability (DoS vectors like clock hijacking), and authorization (unauthorized privileges via configuration errors).

    \item Malicious or compromised vendors can insert backdoors or weak security controls into RAN nodes, enabling data exfiltration or subtle traffic manipulation. Covert attacks, such as those targeting the Open Fronthaul (O-FH) interface, demonstrate the feasibility of stealthy service disruption.

    \item The O-RAN Alliance has responded by mandating ZT principles, including mTLS for control and management interfaces, RBAC on O1/O2, and secure boot and Software Bill of Materials (SBOM) disclosure to mitigate supply-chain risks.

    \item A holistic strategy requires securing all O-RAN interfaces and enforcing  strong tenant isolation in multi-tenant environments to prevent one tenant from accessing or disrupting another's resources. 
    
    \item Continuous monitoring of RAN components, including anomaly detection on inter-component messages, is critical for identifying malicious or misbehaving procured components.
\end{ZTATakeaways}

\subsection{\MakeUppercase{O-RAN Monitoring and Management}}

When deploying plug-and-play modules, especially those procured from external vendors, it is crucial to understand the extent of control and visibility these applications have within the O-RAN infrastructure. This involves identifying the specific RAN (i.e., E2) nodes to which an application can register, the information it can access, and the actions it can perform to influence RAN behavior.

The O-RAN Key Performance Measurement (KPM) framework, specifically the E2SM-KPM model, allows xApps to subscribe to a broad spectrum of RAN measurements and events, enabling them to manage various use cases. Some examples include:

    \noindent$\bullet$ \textit{\underline{UE Mobility~\cite{oran2025e2smkpm}:}} xApps can monitor per-UE measurement reports from the RIC, such as received signal quality (Reference Signal Received Power/Quality - RSRP/RSRQ) on a given cell. They can also gather QoS metrics specific to each UE, including throughput, delay, data volume, and session activity times. These metrics help profile the behavior of different UEs. Additional O-RAN measurements, such as uplink and downlink data volumes and packet drop rates, are also accessible to xApps.

    The RAN Control model (E2SM-RC)~\cite{oran-e2sm-rc} defines an E2 Node Information service (Style 3), which provides cell configurations and neighbor relationships. This enables xApps to receive real-time updates on changes to a cell’s neighbor list or other serving-cell configurations, which are crucial for handovers and network topology. The UE Information Change service (Style 4) can notify xApps of any changes in UE context information, allowing them to track which UEs are connected, or have been handed over from other cells, including their identifiers.

    \noindent$\bullet$ \textit{\underline{Mobility Control Actions:}} The O-RAN RAN Control service model~\cite{oran-e2sm-rc} offers a Connected Mode Mobility Control procedure (Section 7.6.4) that allows the RIC to command handovers. Using this, an xApp can send a RIC Control message to direct an E2 Node to hand over a specific UE to a target cell. The model supports advanced handover types, such as conditional handover (CHO) or Dual-Active Protocol Stack (DAPS) handover. Additionally, mobility xApps can adjust cell-level mobility parameters via E2, including modifying the handover neighbor allow/block list or adjusting idle-mode mobility settings (e.g., reselection thresholds or priorities) to influence UE roaming behavior. These capabilities are documented in the O-RAN specifications, which include Handover Control messages (Section 8.4.4.1 in \cite{oran-e2sm-rc}) and mention such procedures as controllable via the Near-RT RIC~\cite{oran2025usecases}.

    \noindent$\bullet$ \textit{\underline{QoS-Based Resource Optimization:}} The E2SM-KPM model exposes a comprehensive set of measurements across UE-level, bearer-level, and QoS-flow levels. xApps can subscribe to metrics like per-UE and per-QoS-flow throughput, packet loss/drop rates, latency, and resource usage indicators~\cite{oran-e2sm-rc, oran2025e2smkpm}. The KPM specification also details how 3GPP performance management counters (e.g., PDCP bytes, packet delay) are mapped for E2 reporting. This includes conversions to provide per-UE or per-5QI flow metrics, such as individual UE throughput or delay, rather than only cell-aggregated values. With this, a QoS-optimization xApp can retrieve near-real-time metrics like downlink throughput per UE, packet delay per bearer, block error rates per UE, and scheduler usage per UE.

    \noindent$\bullet$ \textit{\underline{QoS Optimization Actions:}} The O-RAN RAN Control model provides mechanisms to tweak RAN resource allocation and bearer parameters to enforce QoS. In particular, Radio Resource Allocation Control (Control Style 2 in E2SM-RC, Section 7.6.3) lets the RIC adjust various scheduler and link configuration parameters~\cite{oran-e2sm-rc}. For example, an xApp can use this to modify a UE’s DRX cycle (Discontinuous Reception) to balance latency vs. battery usage, control the Scheduling Request (SR) periodicity or Semi-Persistent Scheduling (SPS) configuration for certain traffic flows, and even configure configured-grant periodicity for URLLC services. The xApp may also adjust the CQI (Channel Quality Indicator) reporting table or thresholds to influence how the UE/modulation adapts to link quality. Additionally, O-RAN’s Radio Bearer Control capabilities (part of E2SM-RC) allow xApps to reconfigure data radio bearers (Control Style 1 in E2SM-RC, Section 7.6.2) for instance, changing a bearer’s priority or mapping of QoS flows to radio bearers to ensure high-priority traffic gets the needed resources. In summary, a QoS xApp can dynamically tune scheduling and bearer configuration on the E2 Node (via RIC Control messages) to keep QoS KPIs (throughput, latency) within targets.

    \noindent$\bullet$ \textit{\underline{RAN Slicing:}} Recent O-RAN specifications extend performance monitoring to the slice level. The E2SM-KPM framework~\cite{oran2025e2smkpm} introduces new measurements specific to slicing, allowing xApps to receive periodic reports on slice-specific statistics. For example, an xApp can monitor aggregate physical resource block usage or throughput for a specific slice (identified by S-NSSAI), and track the number of active UEs per slice. O-RAN KPM also supports reporting metrics with the Slice ID (S-NSSAI) as a dimension, enabling per-slice performance monitoring. For instance, a Slice SLA xApp can assess whether the eMBB slice is consuming more than its allocated resources, or whether the URLLC slice is maintaining the expected packet delay within the service-level agreement (SLA). Additionally, the Cell Configuration and Control service model (E2SM-CCC)~\cite{oran2025e2smccc} exposes node- and cell-level configuration data to the RIC. Through E2SM-CCC, an xApp can learn the current slicing configuration of a cell, including the slices configured and their corresponding radio resource allocations.

    \noindent$\bullet$ \textit{\underline{RAN Slicing Actions:}} O-RAN introduced the E2SM-CCC (Cell Configuration and Control)~\cite{oran2025e2smccc} model specifically to support slice resource control. Using E2SM-CCC, an xApp can send a control message to adjust cell and slice configuration parameters in near-real-time. For example, the xApp could modify a slice’s resource quota or priority on a particular cell – effectively telling the E2 Node to allocate more Physical Resource Blocks (PRBs) to a slice or cap a slice’s usage if it’s over-consuming. The E2SM-CCC specification defines JSON-encoded control IEs carrying slice configuration updates (the primary use case being Slice SLA assurance). In practice, an xApp might increase the minimum PRB guarantee for a latency-critical slice during busy hours, or tweak the RAN admission control policy per slice (allowing or blocking new UE connections on a slice) through such control messages. Additionally, the general RAN Control model can also influence slicing: as noted earlier, the E2SM-RC control~\cite{oran-e2sm-rc} style for Radio Resource Allocation even allows adjustment of ``slice level PRB quota'' (Section 6.2.3) on the fly. This provides a more dynamic way to enforce slice resource shares (as a policy). Together, these tools let the xApp actively manage slice resources.


In alignment with the previously discussed use cases, rApps deployed in the Non-RT RIC can monitor a wide range of performance indicators and issue high-level control actions through O-RAN’s open interfaces~\cite{o-ran-non-rt-arch}. Operating on non-real-time timescales, rApps focus on optimizing the RAN infrastructure by leveraging historical and near-real-time data gathered via the O1 and A1 interfaces. These applications have visibility into both user-centric metrics (e.g., UE QoE) and infrastructure-centric metrics (e.g., cell utilization, slice performance). Below, we describe several representative use cases enabled by rApps:

    \noindent(1) \textit{\underline{UE Performance Metrics:}}
    rApps can retrieve per-UE metrics such as throughput, delay, and QoS performance using O1 performance measurement (PM) services from O-DUs and O-CUs~\cite{o-ran-O1-oper-maint}. These metrics allow rApps to identify users who are not meeting service-level expectations.

    \noindent\textit{Associated Action – A1 Load-Balancing Policies:}
    Using these insights, rApps can formulate traffic steering policies that define prioritized target cell lists for specific UEs. These A1 policies are pushed to the Near-RT RIC, which enforces the appropriate handover or dual connectivity adjustments at the RAN level.

    \noindent(2) \textit{\underline{Cell Load and Resource Utilization:}}
    At the cell level, rApps can monitor KPIs such as PRB utilization, cell throughput, and connected UE count using O1 PM data~\cite{oran2025usecases}. These metrics help identify congested cells and available neighboring resources.

    \noindent\textit{Associated Action – A1 Load-Balancing Policies:}
    Based on detected congestion, rApps may update or refine A1 policies to steer traffic away from overloaded cells and toward underutilized ones~\cite{o-ran-use-case-analysis-report}, enabling dynamic load balancing across the RAN.

    \noindent(3) \textit{\underline{Enrichment Information:}}
    Contextual data can be provided via A1 Enrichment Information (EI) services. For example, radio fingerprint maps (mapping UE measurements to likely target-cell signal quality), UE trajectory and mobility profiles (speed, direction), service type (delay-sensitive, etc.), and traffic patterns. The Non-RT RIC derives such data from historical analytics and makes it available to assist decision-making.

    \noindent\textit{Associated Action – A1 EI:} 
    As a complement to direct policies, the rApp supplies A1 enrichment information (e.g., predicted coverage maps or UE mobility patterns) to the Near-RT RIC. This is an action where the rApp pushes contextual data that the Near-RT RIC’s algorithms can use to optimize handover decisions (for instance, pre-emptively switching a UE to a cell that it will physically enter soon).

    \noindent(4) \textit{\underline{Application-level QoE Metrics:}}
    rApps can ingest application-layer QoE (quality of experience) indicators such as video stall duration, latency/jitter, or MOS (Mean Opinion Score) for a service. These can be obtained via an external analytics function or an O-RAN RIC analytics service. By ingesting QoE indicators, the rApp can gauge user-perceived quality that raw network KPIs might not reveal.

    \noindent\textit{Associated Action – A1 QoS/QoE Policy Instructions:}
    The rApp sends high-level policies to guide RAN behavior for better QoE. For example, an A1 policy might instruct the Near-RT RIC to ensure a certain bitrate or low-latency for a group of video users (possibly by prioritizing their traffic or scheduling). 
    While standard A1 policy types cover load balancing and slicing~\cite{o-ran-a1-type-definitions}, custom or vendor-specific intent policies may also be applied for fine-tuned QoE control.

    \noindent(5) \textit{\underline{Network-Level Performance Measurements:}}
    rApps collect radio and transport-layer KPIs such as 
    cell throughput, PRB utilization, RSRP/RSRQ, packet loss – retrieved via the O1 interface from O-CUs/O-DUs (see Figure~\ref{fig:oran-architecture}). Fine-grained PM data (with high granularity in time/location) is used so that the rApp can detect where and when QoE drops (for instance, a cell’s downlink throughput per user). These are standard O-RAN operation, administration and maintenance counters (many aligning with 3GPP metrics).

    \noindent\textit{Associated Action – Dynamic RAN Configuration (via O1):} Using the O1 interface, the rApp can directly tune RAN parameters to improve QoE. For example, it might adjust a cell’s scheduling weights or modify QoS profiles (e.g., allocate more PRBs to a video streaming QoS flow) in the O-CU/O-DU configuration. Such actions are taken if the rApp’s analysis indicates that a configuration change (rather than just load balancing) is needed to meet QoE targets.

    \noindent(6) \textit{\underline{Slice-Level SLA Monitoring:}} 
    The rApp can retrieve per-slice (NSSI) performance measurements via the O1 interface. Examples include: Downlink/Uplink PRB usage per slice, average user throughput per slice, number of active UE or PDU sessions on the slice, and success/failure counts for session setups. These metrics (aligned with 3GPP TS 28.552 counters~\cite{etsi128552}) indicate if a slice is nearing its capacity or failing to meet demand.

    \noindent\textit{Associated Action – Slice Configuration Tuning (via O1):} The rApp reconfigures slice attributes through the O1 interface to add or reduce resources for that slice. For example, it can adjust the slice’s bandwidth share or priority on a given cell, change scheduling policy (to guarantee more PRBs to a URLLC slice vs. an eMBB slice), or modify the slice’s QoS parameters. These configuration parameters are defined in O-RAN’s O1 models (many map to 3GPP slice config in TS 28.541~\cite{3gpp-ts-28.541}) and allow on-the-fly SLA enforcement.

    \noindent
    \noindent(7) \textit{\underline{Beam Performance Analytics:}}
    The rApp can gather detailed measurements for each beam or beam group. Examples include beam-specific Signal-to-Interference-plus-Noise Ratio (SINR) or RSRP, reference signal measurements per beam, user throughput per beam, and beam utilization. These can be obtained via O-RAN E2 Service Models (e.g., E2SM for Beamforming) or vendor-specific O1 telemetry from the O-DU/O-RU~\cite{etsi-tr-104037}. Such data shows which beams serve many users, which beams have poor quality, etc.

    \noindent\textit{Associated Action – Beam Configuration Optimization:} 
    The rApp may recommend beamforming configuration changes to adapt to conditions. For instance, it can issue a policy (via A1) or a configuration update that adjusts the number of active beams, their azimuth/elevation (boresight) angles, or beamwidths per cell~\cite{oran2025usecases}. One concrete example is Grid-of-Beams (GoB) optimization: the rApp analyzes coverage and decides which subset of predefined beams should be active. It then directs the RAN to activate or deactivate certain beams to focus the signal where needed~\cite{etsi-tr-104037}. These high-level beam policies are executed by the Near-RT RIC and underlying RAN nodes.

Based on the various use cases discussed above, it is evident that network applications (dApps, xApps, and rApps) offer a broad spectrum of functionalities, ranging from passive monitoring (e.g., querying KPI metrics) to executing control actions that directly influence RAN behavior and user-plane traffic. These applications can access data related not only to network infrastructure (such as load status and throughput of specific cells) but also to individual user equipment (UE), including transmission behavior and mobility patterns. As such, applications deployed in the RICs have extensive visibility into both the RAN infrastructure and the user plane.

From a security standpoint, especially when embracing the ZT paradigm for O-RAN, the privileges granted to applications and embedded ML components must be strictly governed, particularly in light of the fact that many of these applications may originate from third-party providers. 

To support ZT enforcement in O-RAN, we identify the following essential access control requirements, each of which also presents open research challenges:

\noindent\textbf{(1) Regulating Application Access to RIC Domains.}
Upon deployment, each application must undergo a registration process with its corresponding RIC (Near-RT or Non-RT). This process must be extended to include authorization checks based on O-RAN operator-defined policies. These policies dictate whether an application is permitted to operate within a particular RIC and what capabilities it may be granted. Access should not be implicitly granted based on deployment location; rather, access must be explicitly authorized according to operator-defined trust boundaries.

\noindent\textbf{(2) Regulating Application Access to RAN Nodes.}
Each RIC, whether near-real-time or non-real-time, serves as an authoritative entity that manages a subset of RAN nodes (e.g., O-CUs, O-DUs, O-RUs) or other RICs in a hierarchical and distributed control plane. A critical security control is the ability to restrict which RAN nodes a given application can subscribe to or control via standardized interfaces (e.g., E2, A1, O1). For example, consider a load-balancing xApp deployed in one Near-RT RIC instance. If this RIC manages multiple E2 nodes, then the policy enforcement mechanism must ensure that the xApp only interacts (either passively or actively) with an approved subset of those nodes. This is crucial in real deployments where multiple applications with overlapping functionalities are responsible for disjoint RAN segments. Unauthorized overlaps could lead to conflicting or redundant control actions.

\noindent\textbf{(3) Regulating Granular Access Within RAN Nodes.}
Finally, a key requirement is the ability to constrain which operations an authorized application may perform on a RAN node. This includes restricting read access to specific telemetry (e.g., KPIs, measurements) and write access to configuration parameters or procedures. The challenge lies in the vast surface of control: O-RAN specifications define hundreds of parameters, procedures, and data structures across various E2 service models and O1 management schemas. Even with attribute-based (ABAC) or role-based access control (RBAC) models, fine-grained enforcement is non-trivial. 

One promising idea to address this complexity is to introduce the notion of task scopes or intents as access control abstractions. A task defines a bounded operational objective, encapsulating the minimum set of parameters required to perform it. For example, a task such as ``collect performance metrics for a cell'' would grant read access to parameters, like SINR, PRB allocation, and error rates. O-RAN specifications can serve as the foundation for defining such tasks. For instance, a task related to configuring the QoS of a Data Radio Bearer (DRB) involves 22 parameters, as defined in Section 8.4.2 of the E2SM-RC specification~\cite{oran-e2sm-rc}.

These tasks can be organized into hierarchies, reflecting higher-level intents. For example, the broader intent of ``controlling UE access to a cell'' would involve sub-tasks such as DRB QoS configuration and radio access control~\cite{oran-e2sm-rc}. Such a hierarchy can be represented as a permission tree, where each node (task or sub-task) encapsulates specific read/write privileges. An application can then be granted access to a particular subtree corresponding to its authorized operational scope. This approach enables minimum-permission provisioning: an application is granted only the task permissions necessary to fulfill its role, and no more.
 
Permission management over time presents a significant challenge in the context of dynamic and rapidly evolving systems such as cellular networks. O-RAN specifications are frequently updated to address new security concerns, incorporate additional use cases, or refine existing interface behaviors. Simultaneously, O-RAN applications, whether xApps or rApps, undergo continuous development and may be frequently upgraded to support enhanced functionalities, bug fixes, or new control strategies.

In such an environment, statically assigned permissions can quickly become outdated, misaligned with application behavior, or overly permissive as the underlying specifications evolve. Therefore, an important research direction is the automation of permission management, leveraging formal specification analysis and semantic understanding of application behavior. For instance, machine-readable representations of O-RAN specifications (e.g., interface models, E2SM definitions, O1 YANG models) could be parsed and mapped to application functionality descriptors. This mapping would enable automated reasoning engines to identify which permissions are required—or no longer needed—for a given application version, and enforce the principle of least privilege dynamically.

Such an approach would reduce operator overhead, improve the accuracy of policy enforcement, and adapt to evolving application or specification boundaries. This is especially relevant in ZTA, where any deviation between granted permissions and intended functionality can become a security liability.

\begin{ZTATakeaways}
    \item Access to RIC domains should not be implicitly granted based on deployment location; it must be explicitly authorized according to operator-defined trust policies during registration.

    \item A critical security control is the ability to restrict which RAN nodes (e.g., E2 nodes) a given application can subscribe to or control to prevent unauthorized overlaps and conflicting control actions in distributed deployments.

    \item Fine-grained enforcement of read/write access within RAN nodes is challenging due to the hundreds of parameters defined by O-RAN specifications. A promising solution is to introduce task scopes or intents as access control abstractions.

    \item Due to the dynamic nature of O-RAN (frequent specification updates and application upgrades) statically assigned permissions quickly become outdated or overly permissive. Future research must focus on the automation of permission management by leveraging machine-readable O-RAN specifications (e.g., YANG/E2SM models) to dynamically identify and enforce required permissions.
\end{ZTATakeaways}

\subsection{\MakeUppercase{Related Work}}
Atalay et al.~\cite{atalay2023securing} examine security challenges in third-party xApp deployments, focusing on service-chained, microservice-style architectures where xApps consume services from other xApps rather than reimplementing functionality.
They propose the xApp Repository Function Framework (XRF), a framework for evaluating xApp authentication, authorization, and discovery. XRF is split between clients and a centralized server. The XRF Server, deployable at the edge for lower latency or centrally for wider domain coverage, acts as the authorization server, discovery database, and policy enforcer. The XRF client is co-located with each xApp and functions as a ``middle-man proxy'' for xApp-to-xApp communication and potentially xApp-to-near-RT RIC communication, though supporting the latter would require modifications to near-RT RIC operations and updates to the standards.

When an xApp is instantiated, its co-located XRF client generates a profile including a unique identifier, offered services, deployment location, and load and registers it with the XRF server via mutual authentication. This populates the server’s repository for later discovery and authorization. When a consumer xApp needs another’s functionality, its XRF client issues a discovery request; the XRF server returns matching provider profiles, from which the requester selects one based on criteria such as availability or load. To obtain access, the requester’s XRF client sends an authorization request, and the XRF server issues a signed JWT defining the permitted scope. The requester attaches this token to subsequent API calls, and the provider’s XRF client validates it either locally or via the server’s introspection service before granting access. However, the paper does not specify how security policies are defined or enforced, nor does it evaluate the authentication mechanisms beyond certificate-based authentication, which is necessary but insufficient from a ZT perspective.

El Houda et al.~\cite{elHouda24} propose TrustORAN, a blockchain-based decentralized ZT framework that uses smart contracts and cryptographic mechanisms to authenticate and authorize xApps in O-RAN. TrustORAN employs a two-stage verification process to ensure that only legitimate xApps can access protected resources, mitigating risks from malicious or unauthorized applications.
Vendors deploy Access Control Contracts (ACCs) on the blockchain, each acting as an immutable smart contract governing access to a specific xApp. ACCs encode fine-grained policies as tuples describing the subject, object, protected resources, permitted actions, validity period, and associated access token. These policies are transparent and tamper-proof by design.
When a control xApp (CxApp) requests access to a protected xApp (PxApp), TrustORAN generates an authorization token via keccak256, embedding the CxApp and PxApp addresses, contract address, timestamp, validity period, and a nonce. The CxApp signs the token with its private key to ensure authenticity. The near-RT RIC’s PEP then validates the token’s lifetime, signature, and compliance with the relevant ACC before granting access. TrustORAN also supports mutual authentication: the PxApp verifies the CxApp’s credentials and returns a signed response, establishing bidirectional trust.
ACCs further enforce security through Solidity modifiers. OnlyOwner restricts sensitive operations (e.g., adding xApps or updating policies) to the vendor, while OnlyxApps ensures that only authorized xApps or the owner can query or verify access decisions. These mechanisms prevent unauthorized manipulation of access control logic.

Hung et al.~\cite{hung2025anomaly} discuss the problem of security vulnerabilities in the O-RAN's Near-RT RIC, specifically focusing on threats posed by third-party xApps and E2 nodes via the E2 interface. The authors propose an anomaly traffic detector designed to safeguard the Near-RT RIC by verifying signaling legality through a state machine analysis module and checking packet conformance. This detector aims to mitigate DoS attacks and other exploits by identifying and blocking malicious traffic in real-time, demonstrating its effectiveness against known vulnerabilities and O-RAN security specifications.

Eiza et al.~\cite{eiza2025hybrid} propose a hybrid ZT deployment model designed to secure the O-RAN control plane components within 6G networks. The authors propose a novel approach that combines enclave-based segmentation for macro-level isolation with application sandboxing for micro-level isolation of elements such as xApps/rApps. It targets both macro-level segmentation across major O-RAN functional groupings and micro-level isolation of cloud-native applications (notably rApps/xApps), so that only explicitly authorized interactions occur and application code is confined to least-privilege execution contexts. In service of these aims, the system centralizes policy decision-making while distributing enforcement close to the resources and interfaces being protected.

The control logic follows a standard ZT split between policy decision and policy enforcement. A centralized Policy Decision Point (PDP) realized by a Policy Engine (PE) and a PA manages dynamic information-security policies and pushes them outward for enforcement. Enforcement, in turn, is distributed across three elements: enclaves (logical collections of O-RAN resources that share control loops), agents (lightweight PEPs on network elements that forward access requests), and gateways (boundary PEPs that mediate all ingress/egress between enclaves and enforce approved communication paths in coordination with the PA). Policies are centrally administered yet consistently enforced by these agents.

Each enclave groups related O-RAN functions and assigns interface-specific controls to its gateway/agent. In the Management enclave, SMO and the Non-RT RIC are combined; rApps run in a single sandbox, and the gateway enforces policies over A1 (to Near-RT RIC), O1 (to O-DU/O-eNB), and O2 (to O-Cloud). In the Control enclave, the Near-RT RIC, O-CU-CP, O-CU-UP, and O-eNB are colocated; xApps run in a sandbox; the gateway handles E2, Y1, A1, and O1, and departing from a baseline architecture Y1 consumers do not directly access the Near-RT RIC. The Radio enclave (O-DU, O-RU) exposes E2, O1, and Open Fronthaul M-plane at its boundary, while the O-Cloud enclave governs the O2 interface. 

To achieve defense in depth, the authors specify a multi-layered security posture: IPsec at the network layer secures agent-to-gateway transport (confidentiality, integrity, authentication), while mTLS at the application layer provides end-to-end confidentiality, mutual authentication, and integrity between communicating services. This layered composition aligns with ZT principles and is tailored to O-RAN interfaces: for example, A1/O2 use TLS, mTLS, and OAuth, while E2 is mandated to use IPsec for all listed protections. 

Mehrban et al.~\cite{mehrban2025securing} propose a blockchain-based framework to secure the supply chain and lifecycle of O-RAN equipment, directly addressing the vulnerabilities associated with integrating multi-vendor hardware components. The authors leverage a private permissioned blockchain (Quorum) combined with cryptographic firmware authentication to ensure equipment integrity from manufacturing through deployment. During production, each O-RAN device (e.g., O-RU, O-DU) generates a cryptographic hash of its firmware, which is signed using a vendor-specific Hardware Security Module (HSM) and recorded on the immutable ledger. When a device attempts to initialize within the network, the Near-RT RIC triggers an integrity check; smart contracts then automatically validate the device's runtime firmware hash and digital signature against the blockchain registry. Any mismatch triggers an alert via the SMO framework, effectively quarantining the compromised device before it can access network resources. While this approach provides a robust, decentralized mechanism for pre-deployment hardware attestation and effectively mitigates initial supply-chain tampering, its security scope remains limited primarily to the onboarding phase. The framework does not explicitly support continuous, runtime behavioral attestation of the equipment post-deployment to detect subsequent compromises or zero-day exploits. Furthermore, the reliance on a single HSM for firmware signing introduces a centralized point of failure regarding key management, which the authors acknowledge remains an open challenge requiring further research.

Sowjanya et al.~\cite{sowjanya2025abelity} address the limitations of traditional Role-Based Access Control (RBAC) and Transport Layer Security (TLS) in securing the A1 and R1 interfaces within the Non-RT RIC. The authors propose ABElity, a framework that integrates Ciphertext-Policy Attribute-Based Encryption (CP-ABE) to enforce fine-grained, context-aware access control for third-party rApps. In this architecture, sensitive data, such as embedded ML models or optimization policies, is encrypted using a specific attribute policy (e.g., requiring both a specific operational purpose and a geographic region). Consequently, even if a compromised rApp successfully authenticates via mutual TLS and possesses a generally valid role, it cannot decrypt the payload, manipulate policies, or successfully execute a denial-of-service flood unless its specific, cryptographically bound attributes explicitly satisfy the encryption policy conditions.

While this attribute-driven approach significantly hardens the SMO layer against lateral movement and privilege escalation by malicious rApps, it introduces notable operational hurdles. As the authors explicitly acknowledge, ABE imposes considerable computational and network overhead due to its reliance on complex pairing-based cryptographic operations, complicating real-time key distribution in dynamic O-RAN environments. Furthermore, from a strict ZT perspective, the framework focuses heavily on cryptographic access at the point of request but does not detail a mechanism for the real-time revocation of attributes if an rApp's behavioral trust score degrades post-deployment. Without a dynamic feedback loop to revoke keys during runtime anomalies, the system lacks the continuous verification mandate required by a mature ZTA.

%% file: sections/user_plane_zta.tex
\section{\MakeUppercase{Application of ZTA on the Cellular Data Plane}}
\label{sec:up-zta}

ZTA for the cellular data plane focuses on securing the diverse set of communicating entities operating within it, such as smartphones, laptops, autonomous vehicles, robots, as well as on-premises and cloud-based services. Several critical considerations arise in this context. First, according to ZTA principles, only mission-critical communications should be permitted. The nature of communication is tightly coupled with the specific entities involved and the protocol stacks they are allowed to use. Moreover, these entities have varying service requirements. For instance, connections between drones or unmanned vehicle operators and their equipment must maintain extremely low latency to ensure both safety and effective controls. This section explores ZTA directions tailored to such high-priority use cases.

\subsection{\MakeUppercase{Motivating Scenario}}

\begin{figure}[htbp]
\centerline{\includegraphics[width=\columnwidth]{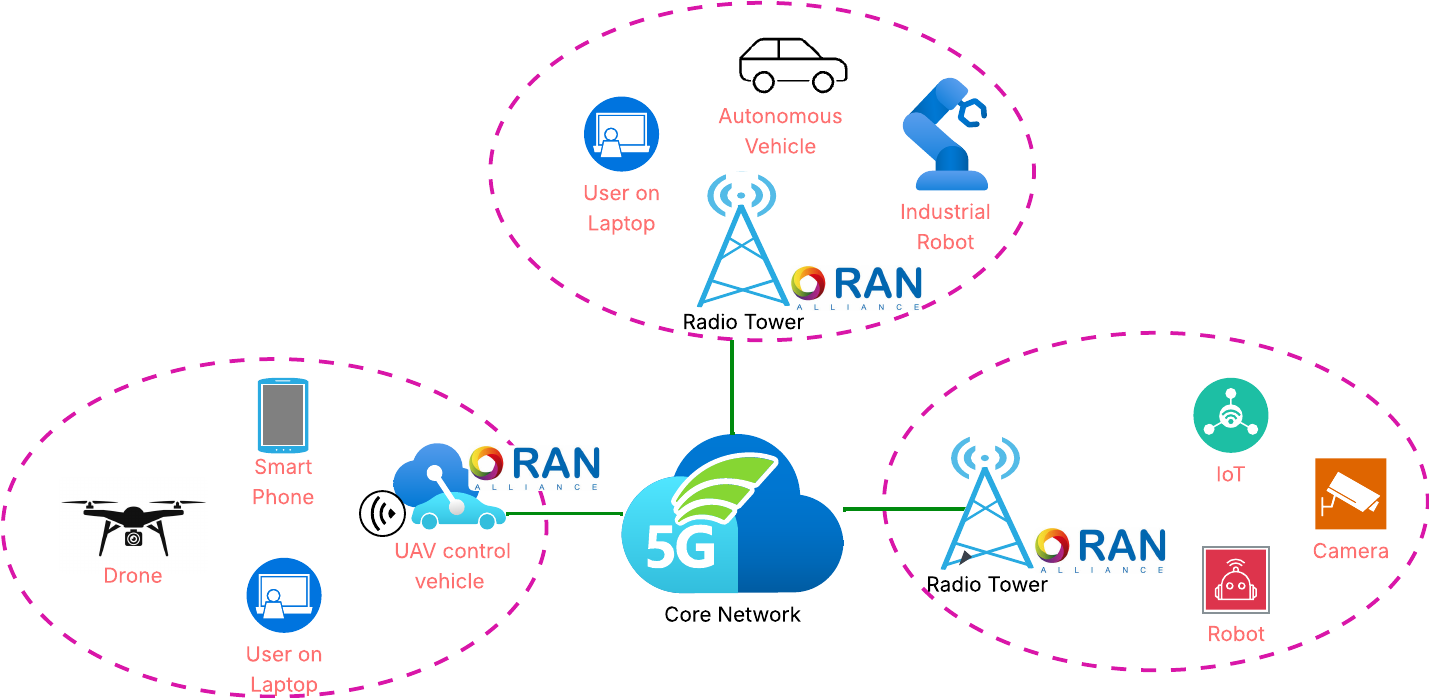}}
\caption{\label{fig:oran-topology-ex} A realistic example of an O-RAN deployment, extending the UAV control vehicle use case~\cite{oran2025usecases}.}
\end{figure}

Figure~\ref{fig:oran-topology-ex} illustrates an extended O-RAN topology in which the unmanned areal vehicle (UAV) control vehicle scenario is embedded within a broader 5G-enabled environment. This extends the use case presented in Section~4.3 of the Use Cases Detailed Specification~\cite{oran2025usecases}. On the left, a UAV, such as a drone, and smartphone, and operator terminal (on a user laptop) connect through the UAV control vehicle, which acts as a mobile edge platform hosting O-RAN components for real-time mission control and high-bandwidth video processing. At the top, autonomous vehicles, industrial robots, and user devices form an industrial/vehicular domain connected via an O-RAN-enabled radio tower. On the right, IoT devices, surveillance cameras, and ground-patrolling robots represent monitoring and machine-to-machine applications, also supported through an O-RAN radio tower. All domains interconnect through the 5G core network, enabling heterogeneous services with differing latency, bandwidth, and reliability requirements. We assume a single O-RAN operator across all three zones, implying shared control/management infrastructure (SMO, Non-RT RIC, Near-RT RIC) and harmonized policy administration across domains.

In this setting, users on the data plane may control applications across domains; for example, a UAV operator can direct the drone from the control vehicle or remotely control a robot attached to a different cell. Mobility introduces further complexity as endpoints such as autonomous vehicles roam between cells, requiring seamless handover between radio towers. While this demonstrates O-RAN’s strengths in software-defined management and dynamic resource allocation, it also highlights the critical need for ZT defenses at the data plane. In a shared, single-operator infrastructure, several threat vectors emerge: unauthorized UEs may attempt to issue commands to mission-specific UAVs; adversaries within a cell may perform lateral reconnaissance by scanning or probing connected devices; malicious traffic may exploit IoT endpoints to pivot across zones; and spoofed control messages may attempt to subvert UAV operations. Some deployments may also include restricted or classified areas, where any attempt by non-authorized devices to connect, even passively, must be denied.

ZT principles require that no communication is trusted by default, even after initial authentication, and that all flows are continuously verified against mission- and domain-specific policies. This includes attestation of device integrity and identity, and evaluation of whether inter-entity communication is permissible under least-privilege constraints. Practically, the Near-RT RIC can monitor E2 telemetry to enforce per-UE access control and anomaly detection in real time (e.g., block out-of-policy UAV control attempts, throttle/blackhole scan-like traffic, isolate compromised IoT nodes), while the Non-RT RIC refines long-term resource and security configurations and distributes updated policies. Additional single-operator use cases include cross-zone policy consistency checks (to prevent policy drift), enforcement of geographic access restrictions (deny devices outside allowed areas), and slice-aware segmentation that prevents lateral movement across zones. In this way, O-RAN can support least-privilege communication, restrict lateral movement, and ensure that mission traffic remains isolated and secure even in multi-tenant, heterogeneous environments managed by one operator.

\subsection{\MakeUppercase{Device Attestation}}

Attestation refers to the process of verifying the integrity and trustworthiness of devices attempting to connect to the network. Within the framework of ZTA, organizations may require connected devices to provide evidence regarding their firmware, operating system, application stack, library versions, and other environment characteristics. This mechanism becomes especially vital in response to newly discovered vulnerabilities. For example, an organization might enforce a policy that denies network access to UE affected by CVE-2023-33106 a critical Qualcomm baseband vulnerability enabling remote code execution through malformed 5G packets.

The concept of remote device attestation has been explored in prior work under various assumptions. For instance, Ambrosin et al.~\cite{ambrosin2016sana} propose a remote attestation framework designed for devices lacking a Trusted Execution Environment (TEE), which is common in resource-constrained IoT devices. Gallagher et al.~\cite{gallagher2022reviewing} discuss the potential for machine learning-based attestation in B5G networks, highlighting system requirements and architectural considerations. However, the application of remote attestation within the O-RAN architecture remains largely unexplored and presents promising directions for future research.

A compelling direction is to implement attestation both at initial network UE attachment and continuously throughout the device’s session lifetime. Neither 3GPP specifications nor O-RAN Alliance documentation currently define UE attestation mechanisms in their use cases, making direct integration of attestation into cellular protocol stacks challenging. As a result, a viable alternative is to design an attestation mechanism at the application layer of the O-RAN control plane. In particular, such functionality could be implemented as a service within O-RAN controllers, such as the non-RT RIC component of the SMO. This approach would require a lightweight UE-side agent capable of collecting relevant attestation data, which could then be securely transmitted to the attestation service within the RIC. One key challenge is that O-RAN applications such as xApps and rApps do not directly communicate with UEs. Instead, their interactions are limited to defined interfaces like E2, through which they influence RAN behavior. To address this issue, an indirect architecture could be employed in which the UE agent communicates with a cloud-based service, which in turn forwards the attestation results to the RIC via an exposed backend API. This design maintains architectural control plane separation while enabling dynamic attestation-driven policy enforcement within the O-RAN ecosystem.

Scalability is a critical requirement for attestation mechanisms. Communications that rely on successful attestation as a prerequisite must be allowed to proceed with minimal latency. To address this requirement, the attestation service must be designed to scale efficiently across large and complex data planes. For instance, environments such as university campuses, research laboratories, military deployments, and large industrial facilities often include a vast amount of connected devices including users' endpoints, sensors, actuators, programmable logic controllers (PLCs), and IoT devices.

\begin{ZTATakeaways}

\item Device integrity attestation is essential to verify firmware, OS, and application trustworthiness before and during network participation.

\item O-RAN and 3GPP specifications currently lack defined UE attestation mechanisms, creating a research gap for integrating attestation into cellular protocol stacks.



\item Scalability and low latency are critical for attestation at data-plane scale in environments with large numbers of heterogeneous devices.

\end{ZTATakeaways}

\subsection{\MakeUppercase{Identity Authentication}}

As discussed in Section~\ref{sec:5g_foundations}, cellular networks only authenticates the subscriber, that is, the SIM card, only. They do not authenticate the identity of the devices or the users using those devices. 
Note that for cellular vendor-locked devices (i.e., where a device does not connect to the network other than the one authorized), this policy is enforced by the device itself and the device is thus not authenticated at the network level. In commercial mobile ecosystems, mobile devices are often ``locked'' to specific network operators, a practice commonly referred to as SIM locking or network locking. This mechanism restricts a device's functionality so that it can only operate with SIM cards issued by a predefined mobile network operator. Importantly, SIM locking is not defined or governed by the 3GPP specifications. It is entirely separate from the authentication procedures standardized in 3GPP, such as 5G-AKA and EAP-AKA, which are used for authenticating the subscription identity (i.e., the SUPI embedded in the USIM) rather than the physical device or its commercial configuration. SIM locking is implemented within the device firmware or bootloader. The mechanism typically checks whether the inserted SIM's Mobile Country Code (MCC) and Mobile Network Code (MNC) match an internal allowlist. If not, the device may refuse to attach to the network or prompt the user to enter an unlock code. Enforcement is performed entirely at the device side, without any interaction with the 3GPP core network functions responsible for subscription authentication or mobility management.

Establishing robust mechanisms and protocols for authenticating both the identity of users and their associated devices is a foundational requirement for securing modern communication systems. Security policies that apply to network endpoints, such as access control based on identity or role, are particularly relevant in this context. For example, in scenarios involving drone communications, access may be restricted to a specific group of authorized users. In such cases, reliable authentication of both the user and the device is essential to determine whether communication should be permitted.

Device authentication methods have been widely studied, with proposed approaches encompassing token-based and certificate-based schemes~\cite{housley2002rfc3280, hummen2013towards}. These methods rely on securely embedded credentials to verify device legitimacy. More recently, research has explored AI-driven continuous authentication techniques, which leverage behavioral characteristics such as network traffic patterns, device motion, or usage behavior to maintain assurance of identity over time~\cite{sodhro2022intelligent, sitova2015hmog}. However, the integration of such continuous authentication mechanisms within the O-RAN architecture remains largely unexplored.

A promising direction for future research is to incorporate identity authentication mechanisms directly within the O-RAN infrastructure. Analogous to remote attestation, such mechanisms could be implemented as dedicated services in the user plane, which would communicate authentication results to the control plane—specifically to the Near-RT RIC. Based on these authentication outcomes, the system could dynamically enforce communication policies, permitting or denying access at a per-UE granularity. This approach aligns with the ZT principle of continuous verification and could enhance access control enforcement within disaggregated RAN environments.

\begin{ZTATakeaways}
    \item In current cellular systems, authentication is limited to the SIM/subscription identity, not the device or user highlighting a structural limitation relative to ZTA goals.

    \item Device and user identity verification are necessary to enforce role- or context-based communication policies (e.g., restricting UAV control to authorized users).

    \item Continuous authentication methods (behavioral or AI-driven) are being researched but are not yet integrated into O-RAN.

    \item A promising ZTA-aligned direction is to implement device/user authentication as a service in the O-RAN infrastructure that reports results to the Near-RT RIC for dynamic, per-UE policy enforcement.
\end{ZTATakeaways}

\subsection{\MakeUppercase{Network Access Control and Stateful Policies}}

NAC policies (Section~\ref{sec:access_control_policies}) are particularly relevant in the context of O-RAN for permitting or blocking traffic associated with individual UEs. For instance, one may wish to allow or block a specific UE from sending traffic to critical systems such as SCADA infrastructure or remotely operated vehicles.

In O-RAN, user-plane traffic traverses from the RAN to the 5G core through the CU-UP (see Figure~\ref{fig:oran-architecture}). This makes the CU-UP a natural candidate for acting as a policy enforcement point, as it mediates all user traffic entering or exiting the radio access network.

However, a key limitation arises from the lack of standardized use cases in O-RAN that support inspection or enforcement based on application-layer semantics (i.e., IP, transport layer and application layer stacks). In contrast to SDN paradigms, where switches may forward packets to a centralized controller for classification (e.g., using Packet-In messages), the O-RAN E2 Application Protocol (E2AP) is limited in scope to RAN control and telemetry functions. Current E2 service models focus on radio resource management, mobility, performance monitoring, and QoS control, but they do not support packet inspection, or application-layer flow identification.

As a result, xApps in the Near-RT RIC are fundamentally constrained to operate on RAN-level metrics such as throughput, CQI, or PRB usage, rather than IP-layer or transport-layer information. While application-layer inspection at the CU-UP may be technically feasible, it is highly implementation-specific and would rely on proprietary vendor capabilities. Furthermore, enforcing such policies would require out-of-band configuration channels not currently defined in the O-RAN specifications.

Traffic filtering and enforcement are typically handled by the UPF in the 5G core, configured via the SMF over the N4 interface. However, in deployment scenarios where the O-RAN infrastructure is operated independently from the core (e.g., a private RAN operator partnering with a commercial 5G provider), the O-RAN operator lacks administrative control over the UPF, SMF, or PCF entities. This separation of control creates challenges for policy enforcement at the edge. An important research direction, therefore, is to investigate how traffic policies can be enforced within the RAN particularly at the CU-UP or Near-RT RIC prior to forwarding traffic to the core network.

Lacava et al.~\cite{lacava2025dapps} made a major step towards these directions. They introduced  dApps, a novel extension to the O-RAN architecture designed to overcome the limitations of existing RICs, such as Near-RT RIC and Non-RT RIC, in handling real-time, user-plane data. dApps are lightweight, plug-and-play microservices co-located with CUs and DUs, enabling sub-10ms control loops and direct access to user-plane data like I/Q samples. The paper proposes a reference architecture for dApps, including a new E3 interface for real-time interaction with RAN nodes and an E2SM-DAPP service model for coordination with xApps. Specifically, dApps are capable of accessing a wide range of user-plane and telemetry data that are generally unavailable to higher-layer O-RAN applications such as xApps and rApps. These include:

    \noindent$\bullet$ \textit{In-phase and Quadrature (I/Q) samples:} Critical for physical-layer security and intelligence tasks such as anomaly detection, spoofing prevention, RF fingerprinting, spectrum sensing, incumbent detection for spectrum sharing, and remote interference analysis. I/Q samples may be collected pre- or post-equalization, from multiple channels (e.g., Physical Uplink Control Channel (PUCCH), Physical Uplink Shared Channel (PUSCH)), and at varying granularities and periodicities. Due to stringent security, privacy, and bandwidth constraints, it is generally infeasible to transfer I/Q data outside the RAN via interfaces such as E2.

    \noindent$\bullet$ \textit{Buffer Status Reports (BSRs):} Generated through interactions between the MAC and RLC layers, BSRs provide real-time insights into buffer occupancy. These may be continuously streamed or retrieved on demand.

    \noindent$\bullet$ \textit{Channel Quality Information (CQI):} Produced by the O-DU, CQI values can be streamed or polled in real time to support adaptive and dynamic resource allocation.

    \noindent$\bullet$ \textit{Channel State Information (CSI):} Similar to CQI, CSI can be streamed or accessed on demand to inform adaptive modulation and coding strategies. dApps further enable dynamic CSI compression in coordination with UEs and the deployment of customized AI/ML models for advanced channel estimation.

    \noindent$\bullet$ \textit{Uplink Sounding Reference Signals (SRS):} Transmitted by the UE and processed by the O-DU, SRS may be continuously provided or requested as needed, offering valuable information for uplink configuration, augmented sensing, and positioning applications.

    \noindent$\bullet$ \textit{Transport blocks and PDUs:} Exposed at multiple protocol layers including MAC, RLC, PDCP, and SDAP according to configurable policies. These may be streamed at defined intervals or retrieved selectively on demand.

    \noindent$\bullet$ \textit{Control-plane information at lower layers:} This includes MAC Downlink Control Information (DCI) and Uplink Control Information (UCI) generated at the DU. Such data can be streamed in conjunction with scheduling decisions or obtained on demand to inform dApp decision-making processes.

    \noindent$\bullet$ \textit{Compute telemetry:} Fine-grained metrics on CPU, memory, and accelerator utilization, which can be streamed at high frequency (hundreds of microseconds) or polled periodically. These measurements enable resource optimization, such as dynamic CPU pinning or energy-state adjustments to improve energy efficiency.

    \noindent$\bullet$ \textit{Fronthaul configuration:} Available on demand to ensure alignment between RAN setup and dApp requirements. This facilitates real-time adaptation of compression strategies to optimize resource utilization.

    \noindent$\bullet$ \textit{Key Performance Measurements (KPMs):} Derived from the Central Unit-Control Plane (CU-CP), Central Unit-User Plane (CU-UP), and DU. While also available through the E2 interface, direct access to KPMs by dApps enables tighter integration with real-time control functions.

Therefore, it is technically feasible to enforce security policies through dApps. This is primarily because dApps are co-located within the RAN nodes, enabling decision-making at sub-10 ms timescales an essential requirement for timely enforcement on user-plane traffic. Moreover, dApps can directly access PDCP (Packet Data Convergence Protocol) PDUs, which encapsulate both user-plane and control-plane information, thereby providing the necessary visibility to implement fine-grained security controls.

Several stateful and context-aware policy enforcement mechanisms have been explored in the SDN literature. For example, Nayak et al.~\cite{nayak2009resonance} introduced the concept of stepwise NAC, where communication privileges evolve based on device state (e.g., internet access is granted only after successful device attestation). Cao et al.~\cite{cao2018cofilter} proposed a stateful packet filter that enforces policies like allowing inbound traffic only if the internal host previously initiated the connection. Kang et al.~\cite{kang2020programmable} developed the Poise system, which uses programmable (P4-enabled) switches to enforce context-aware policies based on environmental variables such as user roles, authentication status, and device type.
Katsis and Bertino~\cite{katsis2025NetSoft} designed an approach to efficiently orchestrate, deploy and manage fine-grained stateful ZT policies in SDN by leveraging programmable data planes. 

Such approaches can be revisited in the O-RAN context, where similar goals may be achieved using programmable RAN components. Specifically, architectures that integrate policy enforcement with the Near-RT RIC offer a promising avenue. The RIC can maintain per-UE state or context information established at attachment and cleared at detachment and use this information to guide xApp action behavior.
xApps are designed to influence RAN operations such as resource allocation, scheduling, and slicing. Hence, policy actions may include reducing a UE’s radio resource allocation in response to suspicious behavior or, in extreme cases, triggering a detach by issuing an RRC Reject. While such actions do not match the fine-grained traffic control possible in core-level firewalls or DPI systems, they provide a viable method for enforcing high-level policies at the network edge.

On the other hand, attacks may originate in the data plane while specifically targeting the cellular control plane. Recently, Wu et al. introduced 5G-SPECTOR~\cite{wen20245g}, an O-RAN–compliant Layer-3 cellular attack detection service that provides a comprehensive framework for identifying control-plane exploits in 5G mobile networks. The framework leverages O-RAN’s programmable architecture and incorporates two principal components to achieve its objectives.
The first component involves telemetry collection at the data plane, where Security Service Module (SECSM) agents are deployed on the O-RAN CUs and DUs. These agents, designed as independent vendor-agnostic plugins, operate in parallel with standard CU and DU processes. By instrumenting the standardized F1 Application Protocol (F1AP) interfaces, the SECSM agents gather both RAN- and UE-related telemetry, including identifiers, bearer lists, and uplink and downlink RRC and NAS traffic. To reduce overhead, RRC and NAS message identifiers are encoded before transmission. The collected telemetry is periodically aggregated into indication messages, and delivered to the Near-RT RIC using E2AP packets over the E2 interface.

The second component focuses on telemetry transformation and attack detection at the control plane. An additional SECSM agent operating at the Near-RT RIC refines the telemetry reported from the data plane into fine-grained MOBIFLOW stream records. This step, which includes state transformation and statistical aggregation, is intentionally separated from the latency-sensitive data plane to avoid performance degradation. MOBIFLOW captures per-UE state transitions and RAN-level statistics, encompassing both temporary and permanent identifiers (e.g., C-RNTI, S-TMSI, IMEI, IMSI/SUCI), security and protocol states, timers, and aggregated counts of connected and idle UEs. These enriched records enable more precise monitoring of control-plane behavior.

The generated MOBIFLOW streams are then consumed by MOBIEXPERT, a programmable xApp hosted on the Near-RT RIC. MOBIEXPERT employs the P-BEST production rule language, allowing operators to define rule sets that evaluate MOBIFLOW records and detect a variety of L3 control-plane attacks. These include BTS resource depletion, blind denial-of-service, IMSI extraction, and null cipher/integrity exploits, which are identified by analyzing abnormal message sequences, state anomalies, or suspicious quantitative patterns. Upon detection, MOBIEXPERT generates alerts and logs, thereby equipping the O-RAN control plane with the ability to perform stateful, near–real-time intrusion detection tailored to cellular protocol exploits.

Beyond detecting control plane exploits targeting the data plane, other recent efforts aim to decentralize the enforcement of UE access policies entirely. Mehrban et al.~\cite{mehrban2025blockchain} propose a multi-layered, decentralized ZT framework designed to secure IoT devices operating within the O-RAN ecosystem. Their architecture is divided into two integrated layers. The first layer employs a combination of Federated Learning (FL) and Transfer Learning (TL) to perform distributed anomaly detection. In this design, an rApp initializes a global model that is subsequently fine-tuned locally by individual IoT devices using their own data; this preserves data privacy and minimizes centralized data collection overhead. The second layer utilizes a blockchain to enforce access control via ACCs. If the threat detection layer identifies anomalous behavior from a specific device, the rApp dynamically updates the ACCs on the decentralized ledger to instantly revoke or restrict that device's network access, creating an immutable audit trail of the intervention.

While this framework successfully combines privacy-preserving machine learning with blockchain-based identity management, an architectural analysis from a strict ZT perspective reveals remaining structural vulnerabilities. 
Primarily, the architecture relies heavily on the rApp to act as the central aggregator and distributor of the global ML model. This establishes the rApp as a high-value centralized target; if the rApp or the underlying SMO layer is compromised, the entire anomaly detection ecosystem becomes vulnerable to top-down model poisoning. Additionally, the framework offloads the computational burden of fine-tuning and running local loss functions directly onto the IoT devices themselves. The authors acknowledge this computational overhead, noting that future work must focus on optimizing the resource efficiency of the framework. In real-world, ultra-low-latency O-RAN deployments, many IoT endpoints (such as headless industrial sensors) are highly resource-constrained and may lack the computational capacity or battery life required to reliably participate in continuous federated training rounds without degrading their primary functions.

\begin{ZTATakeaways}

    \item O-RAN currently lacks standardized mechanisms for application-layer inspection or enforcement; xApps are limited to RAN-level metrics. This limitation motivates research into RAN-level enforcement, including CU-UP and Near-RT RIC-based policy enforcement prior to traffic reaching the 5G core.

    \item dApps offer a feasible path to enforcing security policies within the O-RAN data plane because their co-location within RAN nodes enables sub-10ms decision-making and direct visibility into PDCP PDUs, providing the necessary data for fine-grained security controls.

    \item Existing stateful NAC and programmable data-plane approaches from SDN research can be adapted for RAN-specific enforcement.
\end{ZTATakeaways}

\subsection{\MakeUppercase{State Transfers Across the Cellular RAN System}} 

It is clear that many security policies necessitate the allocation of state, which may include environment variables such as attestation status for UEs. These state variables can be maintained at the Near-RT RIC, since many state-dependent decisions must be made in real time.

During handover, the parameters transferred are those relevant to the CUs at the RAN level, including radio configuration, security keys, and detailed bearer information (e.g., PDU sessions, QoS flow mappings)~\cite{3gpp-ts-38.423}. A significant challenge arises when UEs move between cells that, while belonging to the same administrative domain, are managed by different Near-RT RICs (see Figures~\ref{fig:handover} and~\ref{fig:oran-topology-ex}).

This issue has not been explored in the context of cellular networks. However, similar challenges have been addressed in the context of software-defined networking. For example, Anjum et al.~\cite{anjum2023msnetviews} proposed MSNetViews, a framework for managing enterprise network access control (AC) across distributed geographic locations, which is particularly relevant to organizations operating multiple sites. Their work considers the challenges of user mobility between sites, which requires differentiated access based on location. When a user moves, the federated system must synchronize stateful information to the new site, updating user contexts and policies dynamically. They also analyze scenarios where local policies may vary between sites. Similarly, Gao et al.~\cite{gao2020trident} introduced Trident, a distributed control plane system designed to enhance scalability and efficiency in handling state updates and synchronization across distributed control plane processing nodes. While their work addresses state synchronization in distributed control planes, it does not specifically tackle the challenge of user mobility. In all cases, these problems have not been translated to the context of cellular networks, where mobility requirements are distinct. For instance, autonomous vehicles or drones may move at high speeds within the data plane, necessitating rapid state migration between Near-RT RICs to ensure smooth and uninterrupted communication.

This scenario highlights the need for \textit{inter-Near-RT-RIC handovers}, where application-layer context information must be transferred from one RIC to another. One potential approach is to leverage the Service Management and Orchestration (SMO) framework, which includes the Non-RT RIC. In this design, Near-RT-RICs can synchronize stateful information with the Non-RT-RIC. When a UE migrates from a base station managed by one RIC to another managed by a different RIC, the required context data can be retrieved by the target Near-RT-RIC directly from the Non-RT-RIC. The advantage of this method is that it aligns with the existing O-RAN architecture, as the necessary interfaces already exist to support such data exchanges.

Alternatively, Near-RT-RICs could communicate directly with each other via a dedicated interface, thereby avoiding the additional communication overhead incurred when involving the Non-RT RIC in the process. However, direct inter-Near-RT-RIC communication is not currently supported by the O-RAN specifications.

\noindent\textbf{Considerations.} To address the complexities of high-mobility scenarios (e.g., autonomous vehicles or drones), relying on reactive state synchronization presents a significant distributed systems challenge. If the transfer of the security state lags behind the physical handover, the target Near-RT may prompt full re-authentication due to missing state. This race condition may lead to unacceptable service disruption and re-authentication storms across the network. One possible idea to mitigate this state synchronization latency, O-RAN's intelligence layers can be leveraged to execute predictive, pre-emptive state transfers. Specifically, the Non-RT RIC can utilize historical analytics to derive Enrichment Information (EI), such as UE trajectories, speeds, and mobility profiles, which is then passed to the Near-RT RIC via the A1 interface. By employing predictive AI/ML models to anticipate the UE's target cell before the handover process is formally initiated, the system can proactively push the UE's state variables (e.g., attestation status, access policies) to the anticipated target Near-RT RIC. This proactive synchronization effectively masks the communication overhead of distributed state transfers, ensuring the security context is pre-established and preventing service degradation during cross-RIC handovers.

\begin{ZTATakeaways}
    \item State synchronization between Near-RT RICs is necessary to maintain ZT policies when UEs move between cells managed by different controllers. Two potential approaches are noted: (1) leveraging the SMO/Non-RT RIC for state synchronization; or (2) enabling direct inter-RIC communication (not currently supported in O-RAN specifications).
    
    \item Managing state consistency is crucial for ZTA; the O-RAN challenge of inter-Near-RT-RIC handovers requires rapid state migration (e.g., attestation status) to ensure uninterrupted policy enforcement for highly mobile entities.
    
    \item To mitigate the severe latency and service disruptions of reactive state transfers in high-mobility scenarios, O-RAN's intelligence layers can be leveraged for predictive state synchronization. Utilizing A1 Enrichment Information , the system can proactively push the security context to the anticipated target Near-RT RIC prior to the handover, ensuring continuous ZT enforcement without triggering re-authentication storms.
\end{ZTATakeaways}



\subsection{\MakeUppercase{Preventing Handovers to Restricted Physical Areas}}

Given that RAN deployments often span wide and diverse geographic regions, enforcing security policies that regulate communication at the cell level is essential, particularly in domains such as military networks. In scenarios where certain physical areas fall outside trusted control (e.g., contested or adversary-occupied zones), it becomes critical to restrict access to classified resources. Preventing a UE from establishing or maintaining connections in such regions is a key strategy for safeguarding against potential adversarial exploitation.

The O-RAN specifications acknowledge the need for such enforcement mechanisms as part of their defined use cases and functional requirements. Specifically, the Near-RT RIC supports radio access control as a means to enforce operator-defined access policies~\cite{oran-near-rt-ric-use-cases}. This includes the ability to dynamically influence UE admission decisions at runtime. According to Section 7.6.5.1 of the E2SM-RC specification~\cite{oran-e2sm-rc}, the Near-RT RIC can coordinate with the gNB to reject a UE's connection request by triggering an RRC connection reject message. Such capability allows the operator to enforce access restrictions in real time based on policy conditions, effectively preventing communication attempts from or to undesired geographic regions.

\ZTATakeaway{The Near-RT RIC can reject connection requests via RRC Connection Reject messages to enforce access restrictions in contested or untrusted physical zones. This capability enables operators to dynamically prevent communication from or to geographic areas outside the trusted domain.}

\section{\MakeUppercase{A Framework for Network-wide Security Policy Control}}
\label{sec:proposed-framework}

There is a critical need for a comprehensive framework to support the specification and enforcement of security policies across O-RAN-based cellular networks. Such a framework must facilitate not only the definition of security policies but also their deployment and enforcement across potentially distributed environments comprising multiple RAN nodes and RICs.

\subsection{\MakeUppercase{Security Policy Specification}}

The first essential step in network-wide security policy control is the specification of policies that govern UE network permissions such as determining who can communicate with whom and under what conditions. Over the years, several policy specification languages have been proposed to address this challenge.

One such language is the Next Generation Access Control (NGAC) language developed by the National Institute of Standards and Technology (NIST)~\cite{ferraiolo2016comparison}. NGAC provides a formal, graph-based approach to ABAC, suitable for dynamic and complex environments. It represents access control policies using directed graphs, where nodes capture users, objects, attributes, and policy elements, and edges encode relationships such as assignments, associations, and prohibitions. Katsis et al.~\cite{katsis2021can, katsis2022neutron} introduced the NEUTRON framework, which leverages graph-based abstractions for policy specification. For instance, the communication requirements graph within NEUTRON specifies how various network entities interact (e.g., via protocol stacks) and defines access permissions based on user roles.
Another notable example is Pyretic, proposed by Monsanto et al.~\cite{monsanto2013composing}, which is a high-level, imperative language embedded in Python, tailored for SDN. Pyretic enables modular programming and facilitates the composition of independently defined packet-processing policies over abstract network topologies. This significantly raises the abstraction level for programming OpenFlow-enabled networks.

Looking ahead, there is an important opportunity to design a policy framework specifically tailored to the needs of O-RAN-based architectures. Such a framework should offer expressive capabilities for a broad range of policy types enabling both stateless and stateful enforcement mechanisms. In addition to defining permissible endpoint communications, the framework must support policies that capture QoS requirements, which are especially relevant in cellular networks. For example, communications involving latency-sensitive UEs should be mapped to network slices offering ultra-low latency, while others may be assigned to higher-latency slices. Furthermore, support for location-based policies is essential to dynamically control UE communication based on the geographical point of network attachment.

\subsection{\MakeUppercase{Security Policy Enforcement}}

Within the O-RAN architecture, the Non-RT RIC, located in the SMO layer, serves as the central component for orchestrating network-wide security policies. Based on the specified policies, the Non-RT RIC can configure network slices tailored to various QoS requirements. These configurations are enforced through the O1 interface and may involve the allocation of radio resources to support distinct QoS classes for different communication flows.

Once defined, these policies can be dynamically distributed at runtime to the appropriate Near-RT RICs, each managing a portion of the RAN. Near-RT RICs utilize xApps for real-time monitoring and control of the radio access network. These xApps play a vital role in enforcing security policies on-the-fly. For example, an xApp may be responsible for tracking the attestation status of UEs; based on policy, a UE with an invalid or unverified attestation state may be denied access to the network.
Furthermore, xApps can inspect and regulate user plane traffic via the E2 interface to ensure compliance with policy constraints. If a UE attempts to access unauthorized services or destinations, xApps can intervene by directing the RAN to terminate such sessions.

Given the distributed nature of O-RAN deployments, it is essential to establish an end-to-end policy enforcement pipeline. Such a pipeline should enable the high-level specification of security policies and facilitate their automatic translation and enforcement within the O-RAN ecosystem. This approach ensures that policy-driven decisions made at the SMO level are consistently enforced across all RAN components in a scalable and efficient manner.

\noindent\textbf{Latency Considerations in Policy Enforcement.} While the distributed enforcement of security policies across O-RAN components (such as Near-RT RICs and O-CUs) aligns with ZTA principles, it introduces considerable computational and runtime overhead. The foundational ZTA requirement of continuous verification (which entails ongoing identity checks, policy evaluations, and telemetry lookups) can consume critical timing margins. This overhead is particularly problematic for latency-sensitive network paths and resource-constrained edge workloads, where strict control-loop budgets (e.g., 10 ms to 1 second for the Near-RT RIC) must be maintained to avoid service degradation or self-inflicted DoS caused by overzealous policy checks.

To reconcile the high-assurance requirements of ZTA with the stringent latency constraints of O-RAN, recent literature has proposed adaptive enforcement strategies. For instance, Mehrban et al.~\cite{mehrban2025integrating} proposed the integration of the Risk-Adaptive Access Control (RAdAC) within the O-RAN architecture to mitigate these latency tensions. RAdAC dynamically modulates the depth of verification and authorization as a function of contextual risk, preserving the ``never trust, always verify'' posture without uniformly burdening the network. In this model, Policy Decision Points (PDPs) evaluate environmental state and trust signals continuously. Under low-risk, steady-state conditions, the PEPs maintain a lean enforcement path by relying on local caches, short-lived credentials, and previously attested device posture. Heavy verification processes such as full re-authentication, deep policy evaluation, or immediate token revocation, are triggered only when behavioral drift or elevated threat indicators are detected.

Beyond optimizing the RIC-level control loops, achieving ultra-low-latency policy enforcement requires pushing decision-making even closer to the network traffic. As discussed in Section~\ref{sec:up-zta}, the emerging paradigm of dApps offers a highly effective mechanism for this localized enforcement~\cite{lacava2025dapps}. Co-located directly within the O-DU and O-CU, dApps bypass the latency inherent in traversing the E2 interface to the Near-RT RIC. Because they have direct visibility into user-plane traffic and lower-layer telemetry (such as PDCP PDUs, MAC control information, and raw I/Q samples) dApps can function as highly responsive, decentralized PEPs. This architectural shift allows for instantaneous, stateful security decisions, such as dropping anomalous packets or isolating compromised UEs at sub-10 ms timescales. By delegating real-time enforcement to dApps at the edge of the data plane, the ZTA framework significantly reduces the need for constant RIC consolidation, ensuring that high-throughput, latency-critical operations remain both secure and performant.

Finally, the highly virtualized and cloud-native nature of the O-RAN ecosystem offers another structural solution to the latency challenge: horizontal scaling and workload distribution. Because O-RAN components, such as the Near-RT RIC and O-CU, are typically deployed as virtualized or cloud-native network functions on the O-Cloud, operators can dynamically instantiate multiple instances of these components to distribute the policy enforcement workload. This approach mirrors distributed state management techniques proven in the broader SDN domain. For example, Gao et al.~\cite{gao2020trident} introduced the Trident framework, which utilizes a distributed runtime to manage reactive network policies. By splitting the policy evaluation workload across multiple processing ``shards,'' Trident processes network requests in parallel. This distribution of messages into multiple queues effectively reduces buffering time and latency, significantly boosting overall system throughput. Applying a similar multi-instance deployment strategy within O-RAN would allow the network to handle the heavy computational load of continuous ZTA verification in parallel, preventing single points of congestion and preserving the real-time responsiveness required by the cellular data plane.

\section{\MakeUppercase{ZTA Implications Across O-RAN Deployment Scenarios}}
\label{sec:deployment-implications}

The necessity and implementation strategy for the ZTA mechanisms discussed in the preceding sections vary significantly depending on the specific O-RAN deployment model. The structural trust boundaries of each scenario dictate where policy enforcement must occur and how the RAN interacts with the Core Network.

\noindent\textbf{Hybrid Public-Private Deployments:} In hybrid environments where an enterprise that operates the local RAN but relies on a public MNO for the 5G Core, the enterprise lacks administrative control over the Core's UPF and PCF. Consequently, the enterprise cannot rely on the Core network to enforce localized ZT data-plane policies (e.g., blocking unauthorized IoT devices from accessing a local SCADA server). For these deployments, pushing NAC to the edge, specifically utilizing the CU-UP or localized dApps for sub-10 ms user-plane inspection, is not merely an optimization; it is a structural necessity to maintain the enterprise's isolated trust perimeter.

\noindent\textbf{Multi-Operator RAN Sharing:} When multiple MNOs share the same physical O-RAN infrastructure (e.g., Shared O-RUs), the attack surface expands laterally. In this scenario, strict micro-segmentation and granular task-scoping become the most critical ZTA controls. Because multiple Near-RT RICs from different administrative domains may attempt to optimize the same shared radio resources, a compromised or maliciously procured xApp from Operator A could execute control actions that inadvertently degrade Operator B's service. Therefore, secure procurement pipelines with behavioral validation for AI/ML models and dynamic, machine-readable permission boundaries are paramount to ensure tenant isolation at the radio level.

\noindent\textbf{Traditional Operator-Owned Deployments:} In traditional, macro-scale deployments, a single MNO owns both the public O-RAN infrastructure and the 5G Core. Because the MNO has full administrative control, they can seamlessly enforce data plane policies at the Core (e.g., via the UPF or SMF). However, public networks serve millions of unmanaged, diverse consumer devices, making deep endpoint attestation impractical. Consequently, the RAN edge remains a highly untrusted environment. Under ZTA, the MNO cannot rely solely on the 5G Core as the primary PEP. Malicious control plane actions (e.g., a rogue xApp triggering a localized signaling storm) or physical-layer data-plane attacks (e.g., adversarial I/Q manipulation) are entirely invisible to the UPF, AMF, and other Core functions, which operate strictly at the network and session layers~\cite{3gpp-ts-38.410,3gpp-ts-23.501}. Because the O-RAN architecture delegates autonomous decision-making to the edge, manipulated telemetry will degrade the radio environment before the Core is even aware of an anomaly. Therefore, traditional MNOs must deploy ZTA anomaly detection at the RAN edge to identify and neutralize RAN-specific exploits locally, preventing them from cascading into the Core network.

\noindent\textbf{Private Cellular O-RAN:} Fully private networks (where an enterprise owns a localized RAN and Core) are structurally similar to traditional deployments but are characterized by a vastly different threat model and operational scope. Private O-RANs are often deployed in mission-critical, highly isolated environments (e.g., industrial factory floors or military tactical networks) that rely heavily on URLLC. In these ``air-gapped'' settings, introducing third-party plug-and-play modules (xApps/rApps) creates severe supply-chain risks that could bridge the isolated network to the outside world; thus, stringent secure AI/ML procurement processes are an existential necessity. Furthermore, unlike public MNOs, private networks operate highly controlled, enterprise-owned endpoints (e.g., industrial robots, headless sensors). This allows private operators to strictly enforce robust remote device attestation and continuous identity authentication at the RAN edge, ensuring compromised IoT endpoints are quarantined before they can pivot laterally into the local Core.

%% file: sections/industry.tex
\section{\MakeUppercase{Selected Approaches by Industry}}
\label{sec:industry}

\subsection{\MakeUppercase{Nokia's O-RAN Security Approach}}

Nokia Corporation has released a high-level white paper discussing their approach to securing O-RAN~\cite{nokia_open_ran_security_2025}. The document mainly outlines the challenges of O-RAN security in Nokia's view and revolves around the following key pillars:

    \noindent\textbf{Integration of New Network Functions.} The O-RAN architecture introduces novel functions that expand the potential attack surface. Notably, these include the Non-RT RIC and the SMO framework. The SMO framework is defined by the O-RAN Alliance as an open platform designed to manage multi-supplier, multi-technology networks. Communications Service Providers (CSPs) have the flexibility to select and integrate components for the SMO framework from various suppliers.

    \noindent\textbf{Open Interfaces and Interoperability.} The O-RAN Alliance has defined new open interfaces to support new functionalities, which, however, consequently increase the attack surface and present potential opportunities for threat actors to gain system access. These interfaces include O1 and O2, which are related to the SMO framework, A1 and R1, which are associated with the Non-RT RIC, and the Open Fronthaul interface situated between the Radio Unit (RU) and the Distributed Unit (DU).

    \noindent\textbf{Multi-Supplier Environment.} The O-RAN architecture is designed to accommodate a diverse range of suppliers. While this aims to enhance resilience and diversify the supply chain for critical communications infrastructure, it necessitates ensuring the integrity and authenticity of all assets within the O-RAN system. The integration of components from different suppliers, if not meticulously managed, can lead to unintended vulnerabilities. Nokia advises CSPs to conduct thorough due diligence checks on all suppliers and implement stringent security controls for all third-party network access to mitigate the risks presented by multiple suppliers.

    \noindent\textbf{Supply Chain.}  In a multi-supplier environment, securing the entire supply chain becomes a significant challenge. Reports, including Nokia’s threat intelligence analysis and a report from the European Union Agency for Cybersecurity (ENISA), indicate a rapid increase in supply chain attacks, which are often fueled by geopolitical tensions and swift technological advancements. The supply chain is identified as an efficient route for attackers to gain access to and control target systems. Common attack techniques employed to compromise a supply chain include malware infection, social engineering, brute-force attacks, and the exploitation of software and configuration vulnerabilities, as well as leveraging open-source intelligence (OSINT). The complexity of both technology solutions and the supply chain is heightened in a multi-supplier Open RAN system.        

Nokia's O-RAN security is built upon eight key pillars designed to mitigate risks associated with the expanded threat surface in this evolving architecture. These pillars ensure that security is an integral part of the Open RAN deployment, from design to continuous operation:

    \noindent\textbf{Design for Security (DFSEC).}
    Nokia’s DFSEC process~\cite{nokia_dfsec_2023} integrates security and privacy from the inception of product development through to the end of the lifecycle. It includes defined decision points, vulnerability database maintenance, secure patch management, and adherence to cloud security policies. The use of hardware-based security mechanisms, such as Trusted Platform Module (TPM), ensures integrity and authenticity beyond what software-based controls can offer.

    \noindent\textbf{ZTA Adoption.} Nokia embraces the NIST-defined ZTA, built on the principle of ``never trust, always verify.'' 
    Nokia claims that its implementation aligns with the ZT Maturity Model defined by the US CISA and spans all RAN components, supporting end-to-end security monitoring and real-time threat detection.

    \noindent\textbf{Supply Chain Security.}
    Recognizing the risk posed by complex, multi-layered software and hardware supply chains, Nokia employs the Secure Supply Chain Consumption Framework~\cite{openssf_s2c2f_2022_2024} and conforms to ISO/IEC 5230:2020~\cite{openchain_iso5230_2020}. Software Bill of Materials (SBOMs) and strict vendor selection criteria based on the Open Source Security Foundation (OpenSSF) Scorecards~\cite{openssf_scorecard_2020_2025} are used to ensure traceability, secure sourcing, and integrity throughout the development and delivery pipeline.

    \noindent\textbf{System Integration of Multi-Supplier Solutions.}
    Nokia claims that they apply rigorous integration methodologies and testing procedures to ensure that components from different vendors interoperate securely. This includes disabling unused services, verifying configurations, and enforcing access control policies. Nokia’s dedicated O-RAN Integration Centers in Dallas and Ulm validate multi-supplier solutions under realistic operating conditions.

    \noindent\textbf{Strong Identity and Access Management.}
    Access control is based on multi-factor authentication, RBAC, centralized user management, and privileged user monitoring. The approach is designed to enforce the principle of least privilege and prevent lateral movement. Nokia’s MantaRay NM product is claimed to support seamless integration with external identity management systems and LDAP services.

    \noindent\textbf{Continuous Threat Monitoring and Auditing.}
    The Nokia MantaRay NM platform supports continuous auditing via Audit Trail Logs, which track all configuration changes and security-related events across the RAN. These logs are integrated with the O-RAN operator’s SIEM for long-term analysis. System health monitoring also helps identify anomalies, misconfigurations, or attacks.

    \noindent\textbf{Securing Data in Transit and at Rest.}
    Nokia ensures encryption of data at rest and in transit per 3GPP TS 33.501~\cite{3gpp.33.501}. Public Key Infrastructure (PKI), X.509 certificates, and mTLS secure management-plane communications. 

    \noindent\textbf{Compliance with Security Standards.}
    Nokia’s O-RAN solutions are claimed to conform to standards from 3GPP, O-RAN Alliance, ETSI, NIST, IETF, ISO/IEC, and other key bodies. 
    Nokia is one of the contributing entities to O-RAN security specifications and leads several technical and research groups within the O-RAN Alliance and 3GPP SA WG3. 

This white paper reinforces several key concerns we previously raised in Section~\ref{sec:cp-zta}, particularly around the need for a ZTA, robust access control, and continuous system monitoring. While Nokia’s approach aligns with these principles at a high level, the document stops short of providing technical details on how access control is orchestrated within their systems or how O-RAN system logs are analyzed for anomalies. For instance, it remains unclear which heuristics (if any) are used for detecting attacks or anomalous behavior, and whether AI or machine learning techniques are applied. If such models are in use, further transparency regarding their design, training data, and operational deployment would be essential for evaluating their robustness and effectiveness.

Finally, this document focuses primarily on the security of the cellular control plane. However, as we discuss in Section~\ref{sec:up-zta}, attackers often target user plane assets such as databases, employees, drones, and autonomous vehicles. These systems, which interface directly with users and real-world operations, often represent a more attractive and accessible attack surface. Thus, despite the progress made on securing the control plane, the critical issues related to user plane protection remain largely unaddressed in existing solutions and guidelines, including this one.

\subsection{\MakeUppercase{Ericson's Zero Trust Approach}}

Ericsson has released a white paper~\cite{ericsson2024zta} discussing the critical need for ZTA in securing mobile networks, which are increasingly vulnerable to sophisticated threats. The white paper highlights how traditional perimeter-based security is no longer sufficient and advocates for a ZTA approach that assumes threats exist both externally and internally. The document outlines guidance for MNOs to implement ZTA, aligning with frameworks like NIST 800-207~\cite{stafford2020zero} and the CISA ZTMM~\cite{CISA2023ZTMM}, emphasizing the importance of visibility, analytics, automation, orchestration, and governance. It also details Ericsson's strategy for achieving ZTA, which includes securing products, streamlining deployments, and utilizing the Ericsson Security Manager (ESM) for continuous operational security and compliance. Ultimately, the paper concludes that a dedicated security management function is essential for MNOs to build cyber resilience and meet regulatory requirements in the evolving threat landscape.

Implementing ZTA in mobile networks presents several unique and significant challenges, primarily due to their critical infrastructure status and inherent architectural complexities. As per the white paper, key challenges and considerations for ZTA implementation in mobile networks include:

    \noindent\textbf{Elevated Security Posture Requirement:} Mobile networks, as critical infrastructure, demand a higher security posture compared to typical enterprise networks. Generic ZTA guidance must be adapted to this specialized context.

    \noindent\textbf{Three Distinct Planes:} Mobile networks feature three distinct planes: user, control, and management - each with its own specific contexts, standards, and protocols. 3GPP specifications, while foundational, apply ZTA tenets differently across these planes, prioritizing confidentiality and integrity for the user and control planes, and availability for the management plane, given the severe impact of potential attacks.

    \noindent\textbf{Complexity of Standardized Control Implementation:} Implementing standardized security controls across these diverse planes while simultaneously upholding mobile network service levels is highly complex, necessitating specialized knowledge and specific tools.
    
    \noindent\textbf{Need for Dedicated Security Management:} Beyond existing mobile industry standards, additional security management functionality is crucial for achieving ZTA, particularly to address the demands of security operations. 

    \noindent\textbf{Heterogeneous Technology Stack:} Security-related data, such as configuration and event data, is not limited to the 5G system itself; its type and format depend heavily on specific deployment details, including the various platforms and technologies in use. This necessitates a dedicated security management function capable of bridging the security view across heterogeneous elements, such as physical, virtualized, and containerized Network Functions (NFs) and radios within the mobile network.
    
    \noindent\textbf{Cloud Deployment Operational Challenges:} The use of diverse cloud deployment models (private, public, or hybrid) introduces operational challenges in maintaining a consistent and high-security baseline across different cloud platforms.

    \noindent\textbf{Risk of Manual Configuration Errors:} Manual configuration of network assets and interfaces is time-consuming and increases the risk of misconfigurations and policy conflicts, especially with varied platforms, technologies, and vendors, which is identified as a significant threat.

Ericsson's approach for achieving ZTA in mobile networks is broken down into four key steps:

    \noindent
    \textbf{(1) Secure Approach:} This initial phase involves aligning with regulatory and standards guidelines (such as NIS2 by the European Union~\cite{nis2directive2022}) and shaping a robust security strategy aimed at a ZTA security posture.

    \noindent
    \textbf{(2) Secure Products:} Vendor products, including Ericsson's, are built according to industry standards, laying the groundwork for the implementation of security controls. Ericsson’s Security Reliability Model (SRM) is a comprehensive framework that includes requirements for security functions, controls, and assurance activities, with many requirements mapping directly to NIST ZT tenets and aligning with 3GPP security assurance requirements~\cite{3gpp-ts-33.117}. This ensures products are secure by design.

    \noindent
    \textbf{(3) Secure Deployment:} This step focuses on the proper implementation and configuration of security controls. It emphasizes that automation is critical to enforce micro-perimeter security and enhance scalability and configuration accuracy across the network. Manual security checklists can be replaced by automation tools for configuring the baseline at deployment, mitigating the risk of misconfigurations and policy conflicts. Each NF and its interfaces are implemented with the right security controls and configuration to establish it as a micro-perimeter, with secure-by-default settings automatically verified by the ESM to identify and correct any configuration drift.

    \noindent
    \textbf{(4) Secure Operations:} During ongoing operations, MNOs must ensure security visibility, continuous monitoring, enforcement, and reporting, aligning with NIST tenets on continuously measuring and improving security posture. This step highlights the necessity for additional security management functionality beyond existing mobile standards to meet the demands of security operations.

To specifically address the complex and heterogeneous nature of mobile networks and achieve ZTA maturity, Ericsson emphasizes the crucial role of a dedicated security management function, which is designed to be fulfilled by the ESM. The ESM is designed to tackle operational security challenges and simplify ZTA implementation across diverse technologies and vendors.
Key aspects of Ericsson's approach and how they address challenges include:

    \noindent\textbf{Bridging Heterogeneous Technology Stacks:} Mobile networks involve various platforms (physical, virtualized, containerized NFs, and radios) and diverse cloud deployment models (private, public, or hybrid). The ESM acts as a dedicated security management function that bridges the security view across these heterogeneous elements, ensuring a consistent and high-security baseline.

    \noindent\textbf{Automating Configuration and Monitoring:} To mitigate the risk of manual configuration errors and policy conflicts, a security management function that automates the monitoring of robust security configurations is vital. This ensures a solid security posture and assesses compliance with regional security regulations like NIS2. The ESM facilitates this by automating the onboarding of network assets to monitoring platforms and orchestrating protection and detection activities.

    \noindent\textbf{Delivering CISA ZTMM Cross-Cutting Functions:} Mobile industry standards often do not cover the essential end-to-end aspects of network operations at runtime, such as automation, continuous monitoring and reporting, dynamic policy orchestration, and vulnerability assessment and management. The ESM fills this gap by providing the Visibility and Analytics, Automation and Orchestration, and Governance functions highlighted in the CISA ZTMM. The ESM enables visibility into network security posture by gathering and analyzing network-wide configuration status, logs, and events, both horizontally and vertically across the entire MNO technology stack. This includes continuous monitoring of security configurations and expanding automated log collection and analysis network-wide for centralized, correlated analysis. Additionally, the ESM automates orchestration and response activities, leveraging contextual information from multiple sources to inform decisions.
    Finally, the ESM supports the implementation of network-wide policies for enforcement, with capabilities for dynamic updates and leveraging automation to support enforcement and incorporate contextual information.
    
    \noindent\textbf{Progression through ZTMM Maturity Stages:} Ericsson guides MNOs in incrementally progressing through the ZTMM stages (Traditional, Initial, Advanced, Optimal). The ESM facilitates this journey by enabling:
    \begin{itemize}[leftmargin=*, itemsep=0.1em]
        \item \textit{Initial Stage:} Continuous monitoring and automated log collection and analysis for NFs and underlying cloud infrastructure.
        
        \item \textit{Advanced Stage:} Automated asset discovery, enhanced threat detection capabilities (agent-based and log-based), identification of rogue assets, and establishment of PKI certificate management for NFs.
        
        \item \textit{Optimal Stage:} Comprehensive centralized security visibility and control across all telco nodes, with dynamic monitoring, enforcement, and policy updates, potentially using AI/ML in response to threats or regulatory changes.
    \end{itemize}
    
In essence, Ericsson's approach integrates secure product development (SRM), secure deployment practices with automation, and a robust security management platform (that is, the ESM) to provide a single point of management for ZTA across the end-to-end 5G network. 
Their white paper, however, does not specifically discuss how the Ericson's architectural designs and product developments can be extended and applied to the O-RAN ecosystem.

Ericsson has also published a white paper on O-RAN security discussing the critical need for a ZTA within evolving RAN, especially concerning O-RAN and 5G infrastructure. 
It discusses ZTA Considerations for Specific O-RAN Technologies including:

    \noindent\textbf{(1) Secure AI/ML:} It dives into the security risks introduced by AI/ML in RAN, classifying potential attacks as evasion, poisoning, API, supply chain, and privacy attacks, and links this to the OWASP ML Security Top Ten.

    \noindent\textbf{(2) Secure APIs:} The paper emphasizes APIs as a critical threat vector, ranking them as the number two threat in the cloud by the Cloud Security Alliance, and details industry best practices and OWASP recommendations for securing APIs in Open RAN, such as using HTTPS, strong authentication (PKI X.509, OAuth 2.0), authorization, and input validation.

    \noindent\textbf{(3) Cloud Deployments and Shared Responsibility:} It addresses the new internal threats introduced by third-party infrastructure in cloud deployments for critical infrastructure, including RAN. It highlights the importance of the Cloud Shared Responsibility Model and the need for MNOs to clearly specify security requirements and validate configurations with cloud service providers. It also references the US Enduring Security Framework for protecting 5G Core and RAN in cloud environments~\cite{CISAESF}.

Ericsson's Secure Open RAN Solution is envisioned to address the evolving security challenges in cloud-native O-RAN deployments, aligning with ZTA principles and O-RAN Alliance security specifications. 

The solution's foundation rests on two primary Ericsson products:

\noindent \textbf{Ericsson Cloud RAN:} This software suite implements the goals of cloudification, intelligence/automation, and open internal RAN interfaces for O-RAN, encompassing virtualized DUs and virtualized vCU as Cloud-Native Network Functions running on cloud infrastructure.

\noindent \textbf{Ericsson Intelligent Automation Platform (EIAP):} Functioning as Ericsson's SMO platform, EIAP also delivers the Non-RT RIC and rApps with integrated AI/ML capabilities.

A pivotal element is the established Ericsson SRM which provides a governance structure to embed security throughout the entire product lifecycle, encompassing:  (1) internal software development, (2) consumption of third-party software, including open-source components, (3) secure coding practices, vulnerability scanning, and testing, (4) penetration testing and operational security, and (5) mandatory addressing of OWASP's top 10 risks and the production of a signed SBOM.

\subsection{\MakeUppercase{Discussion}}

It is clear that both Nokia and Ericsson have converging perspectives on the security challenges and design objectives of O-RAN. Their respective white papers discuss the threats introduced by open interfaces, multi-supplier environments, and cloud-native deployments, and they outline both the security mechanisms currently integrated into their products and the incremental path toward ZTA maturity for O-RAN control planes. While these contributions provide a valuable foundation, our work extends the discussion by emphasizing additional control plane considerations that remain underexplored, particularly the governance of access privileges for procured third-party components. Furthermore, we underscore the critical importance of applying ZTA principles not only to the control plane but also to cellular data plane communications, thereby advancing toward a more holistic security posture for the O-RAN ecosystem.

%% file: sections/conclusion.tex
\section{\MakeUppercase{Conclusion}}
\label{sec:conclusion}

The shift toward disaggregated, programmable architectures in 5G and beyond, exemplified by O-RAN, requires a corresponding transformation in security. In this paper, we argued that ZTA must be embedded into the O-RAN ecosystem, where implicit trust is no longer viable. By aligning ZTA principles with O-RAN’s architectural and operational characteristics, we highlighted key challenges in identity management, attestation, access control, and the secure integration of third-party and AI/ML components, and proposed a policy-driven framework that leverages programmable RAN elements for fine-grained, context-aware enforcement.

Despite ongoing standardization, practical ZTA deployment in O-RAN remains limited. Existing specifications provide few mechanisms for continuous trust evaluation, behavioral verification, or secure onboarding. Addressing these gaps is essential to protect both control- and data-plane operations and to support mission-critical, latency-sensitive services. Progress will require scalable policy expression, real-time trust signals across layers, and verifiable AI pipelines that make ZT a built-in property of future cellular networks.

Our survey of industry efforts shows that vendors such as Nokia and Ericsson provide the most detailed public strategies for securing O-RAN. Nokia emphasizes hardware-rooted trust, supply-chain assurance, and DFSEC practices, while Ericsson operationalizes ZTA through its Security Reliability Model, automated deployment pipelines, and Security Manager platform. Both highlight risks stemming from open interfaces, multi-vendor integration, and cloud-native deployments, yet their approaches primarily target the control plane; challenges in securing user-plane assets and AI-driven components remain underexplored.
Our survey builds upon these contributions by identifying unexplored gaps and emphasizing the need for comprehensive, end-to-end ZTA enforcement across the entire O-RAN ecosystem.

Finally, O-RAN’s adoption in commercial networks is still uncertain. Openness introduces innovation and lowers entry barriers but disrupts long-standing business models and supply-chain structures, contributing to cautious industry uptake. The interplay between openness, security assurance, and economic sustainability will ultimately shape the trajectory of O-RAN in future cellular deployments.